\documentclass[
reprint,
superscriptaddress,
amsmath,amssymb,
aps,
pra,
floatfix
]{revtex4-2}

\usepackage[T1]{fontenc}
\usepackage{graphicx}
\usepackage{dcolumn}
\usepackage{bm}
\usepackage{physics}
\usepackage{mathtools}
\usepackage{xcolor}
\usepackage[colorlinks=true, allcolors=magenta]{hyperref}
\usepackage{orcidlink}
\usepackage{soul}
\setstcolor{red}

\usepackage{ulem,xcolor}
\newcommand{\stk}[1]{\ifmmode\text{\textcolor{red}{\sout{\ensuremath{#1}}}}\else\textcolor{red}{\sout{#1}}\fi}

\begin{document}

\title{Perturbative photonic matrix-vector multiplication with reduced phase-shift range}

\author{S. A.\ Fldzhyan\,\orcidlink{0000-0002-9174-8019}}
\email[Corresponding author: ]{fldzhyansa@my.msu.ru}
\affiliation{Faculty of Physics, M.\,V. Lomonosov Moscow State University, Leninskie Gory 1, Moscow, 119991, Russia}

\author{S. S.\ Straupe\,\orcidlink{0000-0001-9810-1958}}
\affiliation{Sber Quantum Technologies Center, Kutuzovski prospect 32, Moscow, 121170, Russia}
\affiliation{Faculty of Physics, M.\,V. Lomonosov Moscow State University, Leninskie Gory 1, Moscow, 119991, Russia}

\author{M. Yu.\ Saygin\,\orcidlink{0000-0001-5494-6801}}
\affiliation{Sber Quantum Technologies Center, Kutuzovski prospect 32, Moscow, 121170, Russia}
\affiliation{Faculty of Physics, M.\,V. Lomonosov Moscow State University, Leninskie Gory 1, Moscow, 119991, Russia}

\begin{abstract}
Programmable photonic meshes provide a promising platform for analog matrix-vector multiplication, but their scalability is often limited by the large phase-shift ranges required in universal interferometer circuits. We introduce a perturbative programming method that operates the circuit near a fixed reference configuration and realizes the target transformation through interferometric subtraction, thereby reducing the required programmable phase excursion. We develop this approach for photonic matrix-vector multiplication architectures based on universal unitary meshes, and low-depth non-unitary constructions based on sums of unitaries. We identify favorable reference configurations through a local conditioning criterion, analyze the phase statistics obtained for random target matrices, and show that perturbative programming produces phase distribution shrinking as the matrix size increases. Identified reference point allows for the subtraction to be carried out electronically after detection, avoiding the interferometric subtraction loss. We further quantify the trade-off between reduced phase range and the intrinsic overhead introduced by the subtraction architecture, and show that for sufficiently lossy phase shifters the reduced phase range can compensate for this penalty. These results identify perturbative programming as a conditional but potentially useful route toward more scalable programmable photonic matrix processors.
\end{abstract}

\date{\today}
\maketitle

\section{Introduction}

Photonic integrated circuits are increasingly being developed as linear processors,  with applications spanning communications, microwave photonics, quantum information processing and  emerging photonic accelerators for linear algebra and machine learning workloads~\cite{Bogaerts2020,Harris2018,Zhou2022,PIC_NNs, tangWaveguidemultiplexedPhotonicMatrix2025, chaiTimeWavelengthmultiplexedPhotonic2026, fldzhyan2026nativeqrfactorizationprogrammable}, and optical switches for datacenters~\cite{SwitchesReview}. A key enabling concept is the programmable photonic circuit, in which a reconfigurable waveguide mesh composed of tunable couplers and phase shifters can be configured post-fabrication to implement a broad class of transfer matrices~\cite{Bogaerts2020,Miller2013}. This capability has catalyzed a shift from application-specific photonic designs toward general-purpose photonic cores supported by electronic control and programming layers.

The development of photonic processors for general-purpose linear algebra is typically associated with the need for large-scale multimode circuits that manipulate many optical channels. Achieving programmability at this scale requires a large number of tunable elements, most commonly phase shifters, which presents a practical obstacle due to footprint, tuning power, and accumulated optical loss~\cite{Bogaerts2020,Harris2018}. These issues are compounded in environments with stringent thermal budgets, where power dissipation associated with control electronics and tuning mechanisms can make even modestly scaled programmable circuits challenging~\cite{Taballione_2021,FSLW_chip,FSLW_Milan}.

A central challenge is that generic linear transforms demand resources that scale unfavorably with dimension. Universal interferometer meshes require a quadratic number of programmable degrees of freedom~\cite{Reck1994,ClementsDesign,Carolan2015}, and conventional Mach--Zehnder interferometer (MZI) meshes typically assume phase shifter modulation ranges spanning $[0,\pi]$ and/or $[0,2\pi]$, independent of circuit size~\cite{ClementsDesign}. This requirement directly impacts footprint and electrical drive requirements, and it exacerbates insertion loss when accumulated phase shifter loss scales with device length~\cite{Harris2014,Qiu2020}.

Three complementary directions can be pursued to improve scalability and compactness. First, advanced material platforms with larger electro-optic and thermo-optic coefficients can reduce the phase shifter length needed for a given modulation range~\cite{BTO_LNOI,BTO_SI_modulator,BTO_RR,STO_Pockels,PZT_cavity,Wang2018,Qi2020}. Second, the design of the programmable elements themselves can be optimized to increase modulation efficiency and bandwidth while reducing footprint~\cite{PZT_cavity,CompactLNOImodulator,Watts2013,Harris2014,Qiu2020}. Third, one may optimize the circuit architecture under explicit assumptions about the target transformation ensemble, potentially yielding phase-efficient alternatives to traditional universal meshes and approaching information-theoretic limits on required tuning ranges~\cite{yuHeavyTailsPruning2023,HamerlyLimitPhases,LowDepthPhotonMNIST}.

In this work we propose an additional, algorithmic approach that targets the required tuning range directly at the level of the implemented linear operation. Specifically, we introduce a perturbative matrix-vector multiplication (MVM) scheme in which the programmable circuit is operated near a fixed reference configuration and the desired transformation is realized as a small deviation from that reference, extracted interferometrically or electronically by subtracting a static replica. This strategy exploits the fact that, for many analog and hybrid analog-digital workflows, multiplying by a scaled matrix can preserve the information content relevant to the computation, while enabling substantial reductions in the required modulation amplitude of programmable elements. Importantly, the perturbative approach is compatible with the three hardware-driven strategies above: it can be combined with improved materials and device designs, and it can be layered on top of phase-efficient unitary-mesh architectures.

We develop this concept for photonic implementations of MVM that require universal unitary interferometers as building blocks, and low-depth constructions for non-unitary operators based on sums of unitaries~\cite{TwoUnitary}. We analyze the trade-off between reduced modulation range and additional insertion loss associated with the subtraction architecture, and identify regimes set by matrix dimension and per-phase shifter loss in which the reduction in programmable-element length can outweigh the overhead loss. At the same time, we emphasize that the practical usefulness of the perturbative regime is limited not only by insertion loss but also by fabrication and calibration accuracy: the static and programmable arms must remain matched well enough that the desired perturbative signal is not obscured by uncontrolled differential errors.

The remainder of the paper is organized as follows. Section~\ref{sec:multiplication} introduces the perturbative MVM principle and its optical implementation. Section~\ref{sec:background} reviews the programmable circuit families considered and establishes the relevant parameterizations. Section~\ref{sec:results} analyzes phase-efficiency improvements for non-unitary targets and presents numerical results, including loss trade-offs. Section~\ref{sec:discussion} summarizes the implications of the perturbative approach, with particular attention to practical limitations and directions for further work.

\section{Main idea}\label{sec:multiplication}

Analog computing typically relies on fast and energy-efficient MVM, which in a photonic co-processor amounts to computing
\begin{equation}
    \mathbf{a}^{(\mathrm{out})}=W\mathbf{a}^{(\mathrm{in})},
    \label{eq:mvm}
\end{equation}
where $W$ is an $M\times N$ matrix and $\mathbf{a}^{(\mathrm{in})}$ is an $N\times 1$ input vector encoded in field amplitudes. More generally, the same hardware is not restricted to a single input vector and can process a batch of $B$ input vectors simultaneously or sequentially, i.e., perform MVM $b^{(\mathrm{out})}=W b^{(\mathrm{in})}$, where $b^{(\mathrm{in})}$ is an $N\times B$ matrix whose columns are the input vectors.

\begin{figure*}[t]
    \centering
    \includegraphics[width=\linewidth]{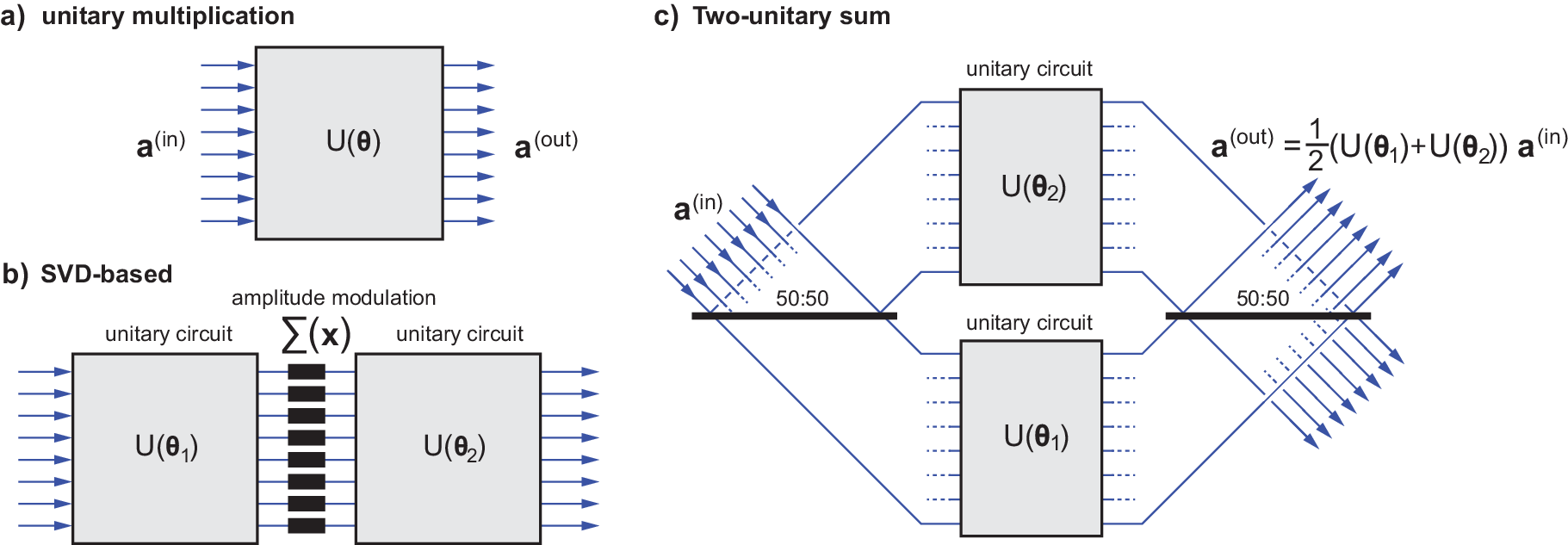}
    \caption{\textbf{Main types of programmable circuits for MVM}: (a) unitary circuit, (b) the standard SVD-based circuit implementing non-unitary MVM, (c) the two-unitary circuit implementing non-unitary MVM.}
    \label{fig:1}
\end{figure*}

The conventional photonic approach to MVM configures the interferometer transfer matrix directly to match the target matrix $W$ and injects amplitudes representing $\mathbf{a}^{(\mathrm{in})}$ into the device. The result is obtained after a single pass through the optical circuit. Fig.~\ref{fig:1} illustrates the main circuit families: (a) a unitary circuit that multiplies the input amplitudes by unitary matrices $U(\boldsymbol{x})$ specified by the parameter set $\boldsymbol{x}$ (e.g., phase shifters)~\cite{Reck1994,ClementsDesign}, (b) the standard SVD-based architecture for general non-unitary matrices~\cite{SVD_Miller}, which exploits a pair of unitary circuits $U_1(\boldsymbol{x}_1)$ and $U_2(\boldsymbol{x}_2)$ and amplitude modulation $\Sigma(\boldsymbol{x})$ implemented, e.g, by absorption modulation~\cite{AbsorbsionModulator}, and (c) the low-depth two-unitary construction~\cite{TwoUnitary}, which exploits a pair of unitary circuits $U_1(\boldsymbol{x}_1)$ and $U_2(\boldsymbol{x}_2)$, but acting in parallel, rather than in sequence as in the SVD-based architecture.

\begin{figure}[b]
    \centering
    \includegraphics[width=\linewidth]{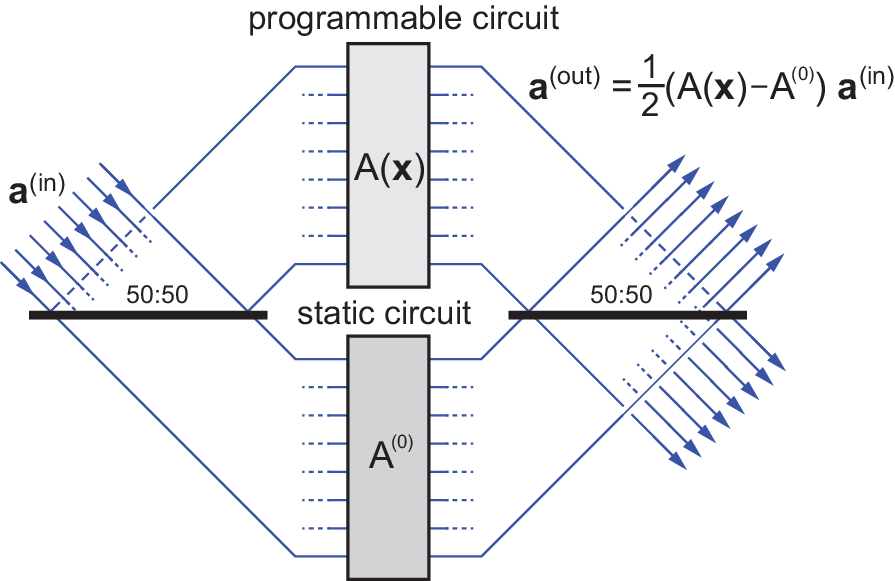}
    \caption{\textbf{Perturbative approach to MVM}: multiplication by a scaled target matrix $\varepsilon W$ through (i) the increment of the circuit transfer matrix $A(\boldsymbol{x})$ relative to a static reference matrix $A^{(0)}=A(\boldsymbol{x}=\boldsymbol{x}^{(0)})$ by properly designing the circuit for its parameters to vary around $\boldsymbol{x}^{(0)}$, and (ii) subtraction of $A^{(0)}$ using interference on balanced splitters/combiners.}
    \label{fig:2}
\end{figure}

Our method introduces a significant deviation from this conventional approach. Instead of implementing the target matrix directly, we use two multimode circuits, as shown in Fig.~\ref{fig:2}. The first is a programmable circuit with transfer matrix $A(\boldsymbol{x})$ controlled by the parameter set $\boldsymbol{x}$. The second is a static circuit that realizes a reference transfer matrix $A^{(0)}$.

The input amplitudes are split equally between the programmable and static circuits by balanced splitters, and the resulting output amplitudes are recombined interferometrically. One set of outputs then yields

\begin{equation}\label{eqn:ref_extraction}
    \mathbf{a}^{(\mathrm{out})}=
    \frac{1}{2}\left(A(\boldsymbol{x})-A^{(0)}\right)\mathbf{a}^{(\mathrm{in})}.
\end{equation}

We require that the reference circuit be chosen such that

\begin{equation}
    A^{(0)}=A(\boldsymbol{x}^{(0)}),
\end{equation}

where $\boldsymbol{x}^{(0)}$ is a reference setting of the programmable circuit. The goal is then to realize the target operation by keeping $\boldsymbol{x}-\boldsymbol{x}^{(0)}$ as small as possible, so that the implemented matrix is a small perturbation around the reference point.

The perturbative approach exploits the fact that multiplication by a scaled matrix can be as useful as multiplication by the original matrix in many analog or hybrid analog--digital workflows. We therefore seek an operating regime in which
\begin{equation}
    A(\boldsymbol{x})=A^{(0)}+\varepsilon\,W(\boldsymbol{x}-\boldsymbol{x}^{(0)}),
    \label{eqn:perturbative}
\end{equation}
where $\varepsilon\ge 0$ is a scaling parameter quantifying the sensitivity of the circuit to the parameter perturbation and $W$ is the actual matrix response. The design problem is thus shifted from implementing an arbitrary target globally to implementing it locally near a bias point.

We note that the subtraction of the reference contribution need not be performed interferometrically. Whenever the complex output amplitudes are measured --- the programmable circuit may implement $A(\boldsymbol{x})$ directly, and the known reference contribution $A^{(0)}\mathbf{a}^{(\mathrm{in})}$ may be subtracted digitaly after detection. The electronic subtraction variant is particularly attractive when multiplication by $A^{(0)}$ is computationally cheap, which, as shown in Sec.~\ref{sec:results}, is exactly the case for the 3-MZI architecture.

\section{Background works}\label{sec:background}

\subsection{Two-unitary design}

As a specific programmable architecture for  non-unitary matrices $A(\boldsymbol{x})$, we consider the two-unitary design depicted in Fig.~\ref{fig:1}(c)~\cite{TwoUnitary}. This construction stands out for its low depth, which is approximately half that of canonical SVD-based circuits, depicted in Fig.~\ref{fig:1}(b)~\cite{SVD_Miller}. Both the designs consist of programmable circuits implementing unitary matrices $U(\boldsymbol{x})$ programmed by parameters $\boldsymbol{x}$ (see Fig.~\ref{fig:1}(a)). It exploits the decomposition of a non-unitary matrix into the sum of two unitaries,

\begin{equation}
    A(\boldsymbol{x})=\frac{1}{2}\left(U_1(\boldsymbol{x}_1)+U_2(\boldsymbol{x}_2)\right),
    \label{eq:twoU}
\end{equation}

where the parameters $\boldsymbol{x}=(\boldsymbol{x}_1,\boldsymbol{x}_2)$ are split into two sets that specify the unitaries $U_1$ and $U_2$.

Arbitrary non-unitary matrices $A$ can be represented in this form provided that the spectral norm satisfies $\|A\|_2\le 1$, as passive linear circuits do not contain gain elements. In the following we focus on the most general setting in which one seeks to realize Eq.~\eqref{eq:mvm} for arbitrary matrices, necessitating universal unitary building blocks.

\subsection{Programmable unitary meshes}

\begin{figure}[t]
    \centering
    \includegraphics[width=\linewidth]{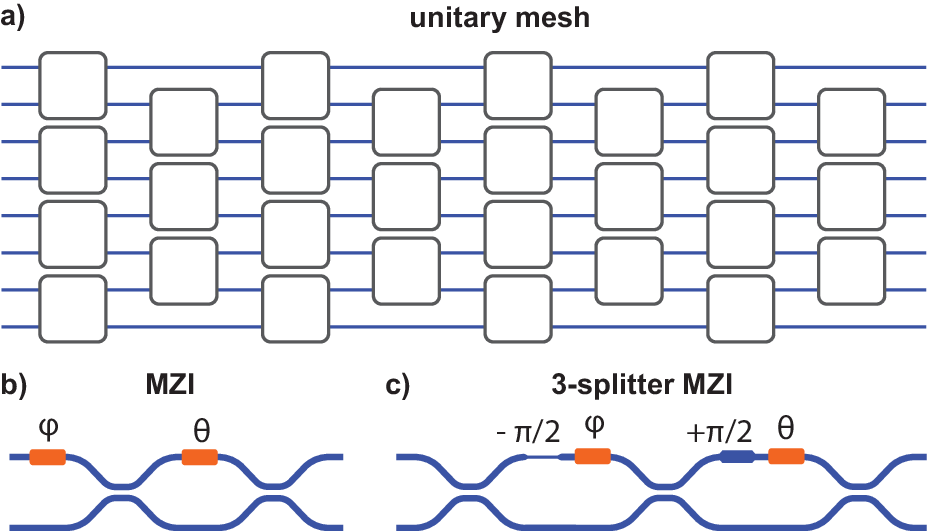}
    \caption{\textbf{Programmable unitary meshes}: (a) the general form of the universal unitary circuit (without one phase shifters layer placed at the input or output modes). The $N$-mode circuit consists of $N$ layers of two-mode blocks. The particular two-mode blocks considered in this work are: (b) the MZI block used in the Clements design~\cite{ClementsDesign}, and (c) the 3-MZI block used in the phase-efficient design~\cite{Hamerly3MZI,HamerlyLimitPhases}.}
    \label{fig:3}
\end{figure}

As particular programmable unitary circuits $U_1$ and $U_2$, we study planar meshes composed of tunable $2\times2$ blocks $B^{(m,n)}(\theta,\varphi)$, shown schematically in Fig.~\ref{fig:3}(a)~\cite{ClementsDesign}. The unitary transfer matrix of an $N$-mode mesh can be written as

\begin{equation}
    U=\varphi^{(\mathrm{out})}\prod_{l=1}^{N}V^{(l)},
    \label{eq:mesh}
\end{equation}
where $V^{(l)}$ is the transfer matrix of the $l$-th layer and has the form
\begin{equation}
    V^{(l)}=\prod_{j\in\Omega_l}T_j^{(l)}\left(\theta_j^{(l)},\varphi_j^{(l)}\right),
\end{equation}
with
\begin{equation}
    T_j^{(l)}(\theta,\varphi)=
    \begin{pmatrix}
        1 & \cdots & & \cdots & 0\\
        \vdots & \ddots & & & \vdots\\
         & & B^{(l,j)}(\theta,\varphi)& & \\
        \vdots & &  & \ddots & \vdots\\
        0 & \cdots &  &\cdots & 1
    \end{pmatrix}.
\end{equation}

It should be noted that, for these circuits to be strictly universal --- that is, capable of implementing MVM with arbitrary unitary matrices --- the mesh must also include a layer of phase shifters outside the $2\times2$ blocks. This additional layer is placed at either the input or the output of the mesh. Together with these input/output phase shifts (not shown in Fig.~\ref{fig:3}(a)), the total number of tunable parameters is $N^2$, matching the number of real degrees of freedom of an arbitrary $N\times N$ unitary matrix.

\subsection{MZI-based unitaries}

The most widely used rectangular universal mesh is the Clements design~\cite{ClementsDesign}, built from MZIs of the form
\begin{eqnarray}
    &B_{\text{MZI}}(\theta,\varphi)=\nonumber\\
    &\frac{1}{2}\left(
    \begin{array}{cc}
        1 & i \\
        i & 1
    \end{array}
    \right)
    \left(
    \begin{array}{cc}
        e^{i\theta} & 0 \\
        0 & 1
    \end{array}
    \right)
    \left(
    \begin{array}{cc}
        1 & i \\
        i & 1
    \end{array}
    \right)
    \left(
    \begin{array}{cc}
        e^{i\varphi} & 0 \\
        0 & 1
    \end{array}
    \right)=\nonumber\\
    &i\left(
    \begin{array}{cc}
       e^{i\varphi}\sin(\theta/2)  & \cos(\theta/2) \\
       e^{i\varphi}\cos(\theta/2)  & -\sin(\theta/2)
    \end{array}
    \right),
\end{eqnarray}
Strict universality requires modulation ranges $0\le \theta\le \pi$ and $-\pi\le \varphi<\pi$, independent of mesh size. This requirement contributes significantly to device length, loss, and drive voltage. More detailed analysis by information-theoretic methods assuming unbiased Haar-uniform distribution has been carried out in~\cite{HamerlyLimitPhases}, indicating more clearly that the phase-shift ranges cannot be narrowed without drop in the transformation quality.

\subsection{3-MZI-based unitaries}

A more phase-efficient alternative was introduced by replacing the canonical MZI block with a 3-splitter MZI (3-MZI)~\cite{Hamerly3MZI,HamerlyLimitPhases}. Using the conventions of Fig.~\ref{fig:3}(c), the block transfer matrix can be written as
\begin{eqnarray}
    &B_{3\text{MZI}}(\theta,\varphi)=\frac{1}{2^{3/2}}\left(
    \begin{array}{cc}
        1 & i \\
        i & 1
    \end{array}
    \right)
    \left(
    \begin{array}{cc}
        ie^{i\theta} & 0 \\
        0 & 1
    \end{array}
    \right)\times\nonumber\\
    &\times\left(
    \begin{array}{cc}
        1 & i \\
        i & 1
    \end{array}
    \right)
    \left(
    \begin{array}{cc}
        -ie^{i\varphi} & 0 \\
        0 & 1
    \end{array}
    \right)
    \left(
    \begin{array}{cc}
        1 & i \\
        i & 1
    \end{array}
    \right)=\nonumber\\
    &
    \frac{e^{i(\theta+\phi)}}{\sqrt{2}}\left(
    \begin{array}{cc}
        \!\sin\frac{\theta-\varphi}{2}+i\sin\frac{\theta+\varphi}{2}\! & \!-\cos\frac{\theta-\varphi}{2}+i\cos\frac{\theta+\varphi}{2}\!  \\
        \!\cos\frac{\theta-\varphi}{2}+i\cos\frac{\theta+\varphi}{2}\!
         & 
         \!\sin\frac{\theta-\varphi}{2}-i\sin\frac{\theta+\varphi}{2}\!
    \end{array}
    \right)\!,\nonumber\\
\end{eqnarray}
where $-\pi/2\le\theta\le\pi/2$, $-\pi\le\varphi<\pi$. This results in asymptotically decreasing phase-shifts $O(1/\sqrt{N})$ with size $N$, because we have tuned the constant  phase shifts

\begin{equation}\label{eqn:3mzi_phases}
    \theta^{(0)}=\frac{\pi}{2},\quad\varphi^{(0)}=-\frac{\pi}{2}.
\end{equation}

Assuming unitary matrices are sampled from a uniform distribution, the authors of \cite{HamerlyLimitPhases} analytically derived bounds for the moments of the phase-shift distribution. In particular, they demonstrated that the maximum phase-shift within this distribution always satisfies the inequality:

\begin{equation}\label{eqn:maxpsi_ineq}
    \text{max}({\boldsymbol{\varphi},\boldsymbol{\theta}})\ge\sqrt{\frac{\pi e^{3/2}}{2N}}\approx\frac{2.65}{\sqrt{N}},
\end{equation}
suggesting that it may be possible to design programmable unitaries with reduced phase-shift modulation ranges without compromising their performance. 

However, the specific examples of MZI and 3-MZI unitary meshes investigated in \cite{HamerlyLimitPhases} unfortunately remain far from approaching the bound described by Eq.~\eqref{eqn:maxpsi_ineq}, as their maximum phase-shift values consistently equal $\pi$ radians. Nevertheless, the 3-MZI meshes exhibit superior phase efficiency in the sense that their phase shift distributions decrease on average as $N$ increases. Specifically, the interquartile range (IQR) of phase-shifts for 3-MZI meshes is:
\begin{equation}
    \text{IQR}^{(3\text{MZI})}\approx\frac{1.9}{\sqrt{N}},
\end{equation}
which is significantly better compared to standard MZI meshes, where the IQR remains constant: 
\begin{equation}
    \text{IQR}^{(\text{MZI})}=\frac{\pi}{2}.
\end{equation}
This makes the 3-MZI architecture a natural starting point for further phase-range reduction by perturbative programming.

\section{Results}\label{sec:results}

\begin{figure*}[hbt!]
    \centering
    \includegraphics[width=\linewidth]{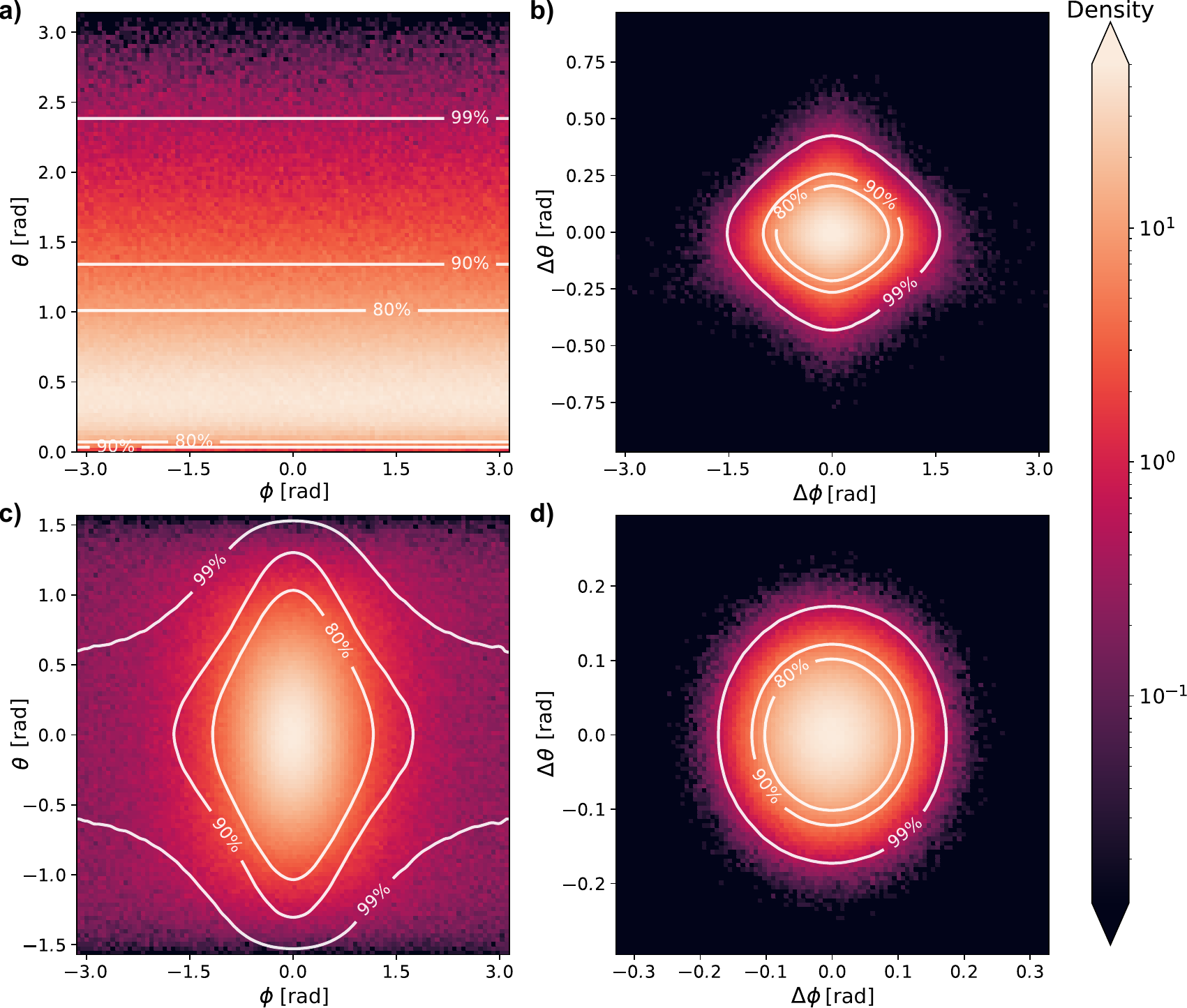}
    \caption{\textbf{Illustration of the perturbative approach for multiplication by dense matrices of size $N=32$}. The plots show the distributions of the angles $\theta$ and $\varphi$ obtained from a two-unitary decomposition of these matrices, contours show the percentage of samples inside. a) MZI decomposition, conventional approach; b) MZI decomposition, perturbative approach with $\varepsilon = \varepsilon_{max}$; c) 3-MZI decomposition, conventional approach; d) 3-MZI decomposition, perturbative approach with $\varepsilon = \varepsilon_{max}$. Each distribution is obtained using $2000$ target matrices. The target matrices are drawn from the Ginibre ensemble and normalized by spectral norm. }
    \label{fig:pretty_N32}
\end{figure*}

\subsection{Choosing the reference matrix $A^{(0)}$}

While the basic idea of perturbative MVM is conceptually straightforward, selecting suitable reference parameters $x^{(0)}$ so that arbitrary matrices $W$ can be realized is nontrivial. In simple architectures whose transfer matrix elements are independently specified by the control parameters, the perturbative method is immediate. A clear example where the perturbative approach works well is in circuit architectures whose transfer matrix elements are independently set by parameters, such as $A_{mn} = x_{mn}$ with $x_{mn} \in [0,B]$, where $B$ denotes the parameter modulation range~\cite{Xbar,WeightBanks}. In such scenarios, the reference parameters can be readily chosen in various ways, for instance, as $A^{(0)} = x_{mn}^{(0)}$, where $x_{mn}^{(0)}$ lies within the allowed modulation range. Consequently, the perturbation term $\varepsilon W$ in~\eqref{eqn:perturbative} can represent any arbitrary matrix, provided that all resulting matrix elements $A_{mn}$ remain within this range.

In programmable unitary meshes, however, the mapping from physical parameters to transfer matrices is strongly coupled. Moreover, it remains unclear whether the perturbative approach is applicable, and if so, under which conditions. We therefore choose the reference parameters by maximizing the local expansion of the volume spanned by accessible perturbations near $\boldsymbol{x}^{(0)}$. This is quantified by the absolute value of the Jacobian determinant:
\begin{equation}\label{eqn:params_condition}
    \boldsymbol{x}^{(0)}=\arg\max_{\boldsymbol{x}}\left|\frac{\partial\text{vec}\left[A(\boldsymbol{x})\right]}{\partial\boldsymbol{x}}\right|,
\end{equation}
where $\text{vec}\left[\cdot\right]$ denotes vectorization of a matrix, $\boldsymbol{x}$ are the set of phase shifts of a circuit.

For 3-MZI meshes, Eq.~\eqref{eqn:params_condition} reproduces the phase-efficient operating point given in Eq.~\eqref{eqn:3mzi_phases}. In this case, the reference matrix $A^{(0)}$ takes the form of an anti-diagonal matrix. This simple structure suggests that the subtraction of $A^{(0)}$ in Eq.~\eqref{eqn:ref_extraction} need not be implemented interferometrically in the optical hardware, but can instead be carried out electronically after detection. Indeed, multiplication by an anti-diagonal matrix is computationally inexpensive compared with multiplication by a dense matrix. As a result, the auxiliary operations required by the perturbative method can be offloaded to an electronic processor, which may significantly simplify the practical implementation.  In addition to simplifying the optical setup, performing the subtraction of the anti-diagonal reference matrix in the electronic domain (subtracting inverted input vectors from the result vectors obtained optically) removes the intrinsic losses imposed in the all-optical perturbative MVM.  

By contrast, for the two-unitary architecture based on standard MZI meshes, applying Eq.~\eqref{eqn:params_condition} leads to substantially more complicated optimal phase settings, whose explicit form is given in Appendix~\ref{app:A0}, and hence a more complicated reference matrix. Unlike in the 3-MZI case, the reference matrix is dense rather than anti-diagonal. As a result, extracting $A^{(0)}$ is no longer a trivial operation and generally requires a full MVM, so the simplification discussed above does not apply in this case.

Importantly, the perturbative approach to MVM with programmable meshes imposes upper bound on the scaling parameter $\varepsilon$ in Eq.~\eqref{eqn:perturbative}. To show this, note that to decompose the matrix $ A = A^{(0)} + \varepsilon W $ into two-unitary form, we must ensure that the maximal singular value (spectral norm) of $ A $ satisfies $ \|A\|_2 \leq 1 $ \cite{TwoUnitary}. Using the triangle inequality, along with the fact that $ \|W\|_2 = 1 $ and the expression for $ A^{(0)} $ in \eqref{app:eq:A0}, we obtain:
\begin{equation}
    \|A^{(0)} + \varepsilon W\|_2 \leq \|A^{(0)}\|_2 + |\varepsilon| \|W\|_2 = \frac{1}{\sqrt{2}} + |\varepsilon|.
\end{equation}
Equality can be achieved by choosing $ W = \mathrm{sign}(\varepsilon) A^{(0)} / \|A^{(0)}\|_2 $. This implies that there is a least upper bound on $ |\varepsilon| $, given by 
\begin{equation}\label{eqn:eps_max}
    |\varepsilon|\le\varepsilon_{\max} = 1 - \frac{1}{\sqrt{2}}\approx{}0.29,
\end{equation}
which, in particular, imposes the minimum level of losses introduced by the approach (see Sec.~\ref{sec:losses} for detailed analysis).

\subsection{Perturbative matrix-vector multiplication}\label{subsec:pert mvm}

We now present the numerical simulations used to characterize the phase distributions for different interferometer architectures and perturbation strengths for multiplication by dense matrices. The target matrices $W$ are sampled from the complex Ginibre ensemble, that is, each matrix entry has independent Gaussian real and imaginary parts with zero mean and variance $1/2$, so that $\mathbb{E}[|W_{ij}|^2]=1$. Each matrix is then normalized to satisfy $\|W\|_2 = 1$, which is the natural normalization for passive optical implementations.

To examine the distribution shapes, for each matrix size $N$ we generate multiple independent target matrices. In particular, for the fixed-size comparison shown in Fig.~\ref{fig:pretty_N32}, we use $2000$ matrices with $N=32$.  Each target matrix is decomposed using two versions of the two-unitary architecture: one based on the standard MZI mesh and one based on the 3-MZI mesh. The corresponding phase-shifts that implement a particular target matrix were obtained analytically, owing to the analytical procedure for the MZI and $3$-MZI meshes~\cite{ClementsDesign,Hamerly3MZI}. For each architecture, we consider both the conventional regime and the perturbative regime with $\varepsilon=\varepsilon_{\max}$.

Fig.~\ref{fig:pretty_N32} shows the full phase distributions for $N=32$ in all four cases. Several features are immediately evident. In the conventional MZI case, the phase parameters occupy essentially the full available range: the $\varphi$ phases are nearly uniformly distributed, whereas the $\theta$ phases are concentrated near zero, in agreement with the known Haar-induced statistics \cite{Pai_2019, Russell_2017}. Passing to the perturbative regime reduces the spread of the phase distribution. This effect is considerably stronger for the 3-MZI architecture, where the perturbative distributions remain centered near the phase-efficient operating point and become substantially narrower. A further reduction is obtained for smaller values of $\varepsilon$; this case is omitted from Fig.~\ref{fig:pretty_N32} for clarity, but is captured in the scaling analysis below.

We next study how the phase distributions evolve with matrix size $N$. As a measure of their spread, we use the interquantile range $\mathrm{IQR}_p$, defined as the width of the central interval containing a fraction $p$ of the distribution:
\begin{equation}
    \mathrm{IQR}_p = q_{\frac{1+p}{2}} - q_{\frac{1-p}{2}},
\end{equation}
where $q_\alpha$ is the $\alpha$-quantile. This quantity generalizes the usual interquartile range, recovered at $p=0.5$. In the limit $p=1$, $\mathrm{IQR}_1$ is twice the maximum absolute deviation and therefore captures the full support of the distribution. In the simulations, we evaluate $\mathrm{IQR}_p$ for
\begin{equation}
    p = 0.5,\ 0.75,\ 0.95,\ 0.99,\ 1.0.
\end{equation}
This makes it possible to distinguish between two qualitatively different scenarios: one in which only the bulk of the distribution contracts while the tails remain broad, and another in which the entire distribution, including its extreme tails, narrows uniformly.

\begin{figure}[hbt!]
    \centering
    \includegraphics[width=\linewidth]{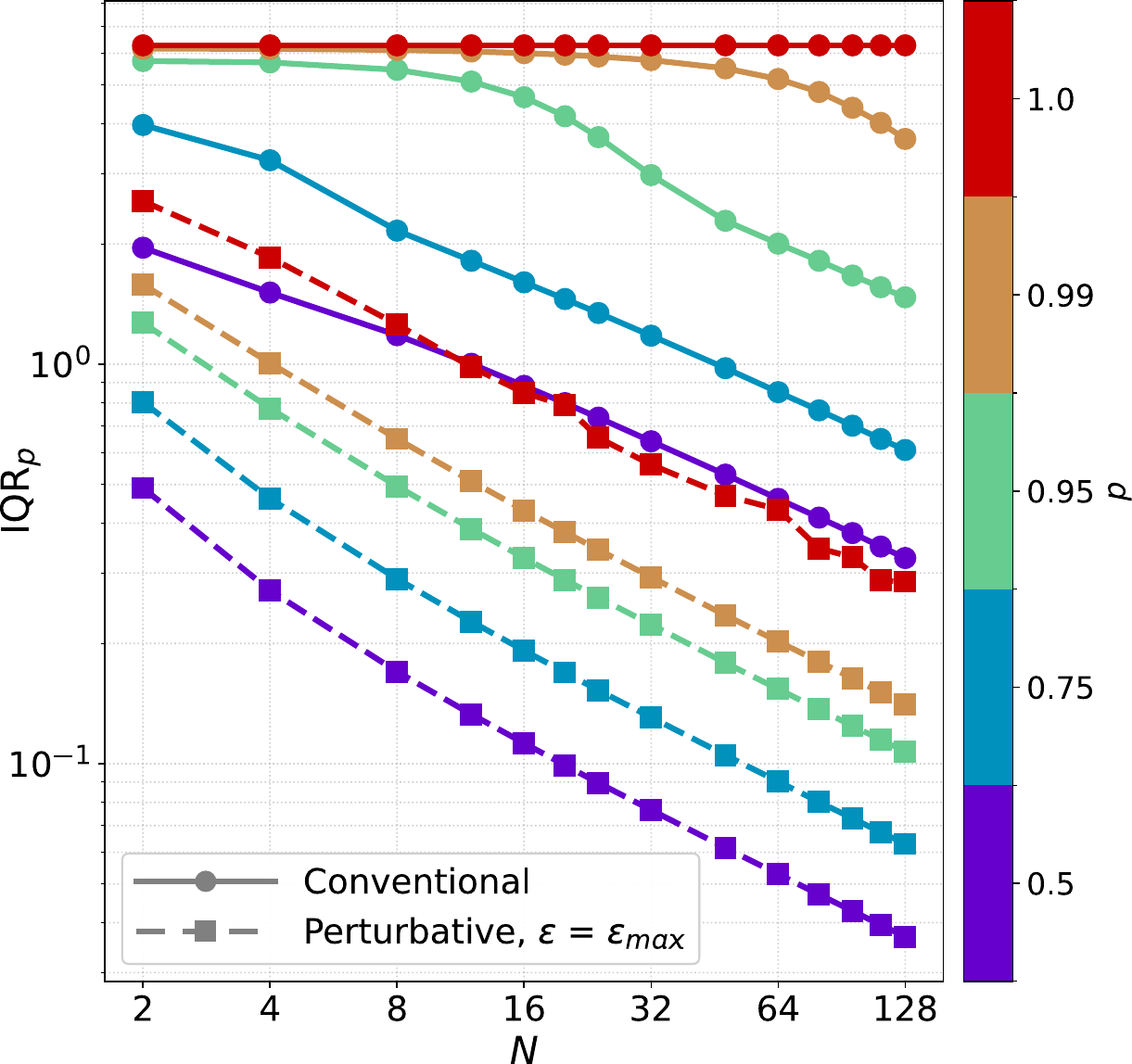}
    \caption{\textbf{Scaling of the phase-distribution width with matrix size in the conventional and perturbative regimes}. Shown is $\mathrm{IQR}_p$ as a function of $N$ for different values of $p$, with the perturbative case taken at $\varepsilon=\varepsilon_{\max}$. The target matrices are sampled from the Ginibre ensemble and normalized by their spectral norm. The number of target matrices has been varied with $N$, for the total number of phase samples matches that obtained from $100$ target matrices of size $N=128$.}
    \label{fig:IQR}
\end{figure}

These trends are summarized in Fig.~\ref{fig:IQR}, which shows $\mathrm{IQR}_p$ as a function of $N$ for the 3-MZI architecture. The main result is the clear difference between the perturbative and conventional regimes. Namely, IQR$_p$ for the conventional regime is substantially larger than for the perturbative: the asymptotic dependencies are related by $\sim\varepsilon$, which is shown analytically in Appendix \ref{appendix:relations}. Moreover, in the conventional case, $\mathrm{IQR}_p$ exhibits the expected $1/\sqrt{N}$ behavior for fixed $p$, but approaches a constant as $p\to 1$. Thus, the central part of the distribution narrows with increasing $N$, whereas the tails remain broad. By contrast, in the perturbative regime the $1/\sqrt{N}$ scaling persists across the entire range of $p$, including the near-extreme quantiles. In other words, the full phase distribution contracts with increasing matrix size.

This behavior is important from the hardware perspective. In the perturbative regime, increasing $N$ reduces the absolute phase range that must be programmed, which can relax the requirements on phase shifter tuning range and control accuracy. An analytical derivation of the $1/\sqrt{N}$ scaling is given in Appendix~\ref{app:sec:entropic}.

Although the analysis above focuses on dense random matrices, the resulting scaling should not be interpreted as universal for all matrix ensembles. Dense MVM is the natural primary target for photonic acceleration, since it is among the most computationally demanding linear-algebra primitives and appears, for example, in neural-network layers and other workloads where a large number of matrix elements contribute to each output. For such high-entropy target ensembles, the perturbative approach leads to a phase distribution whose width decreases with $N$, as confirmed numerically in Fig.~\ref{fig:IQR} and supported by the entropic argument in Appendix~\ref{app:sec:entropic}.

Sparse matrices constitute a different regime. In this case, the target ensemble has much lower entropy, and the phase-range reduction obtained for dense Ginibre matrices need not persist. A limiting example is given by permutation matrices, which are important in photonic switching networks and optical interconnects. As shown in Appendix~\ref{appendix:sparse}, the structure of the Jacobian and the permutation matrices prevents any size-dependent reduction. Lower bound does not produce the same $1/\sqrt{N}$ reduction of the phase range; instead, it gives a size-independent lower bound of order $\varepsilon$. Thus, while perturbative programming is especially favorable for dense high-entropy matrix ensembles, its asymptotic phase-efficiency advantage is ensemble-dependent and is not guaranteed for highly structured sparse transformations.

It is worth emphasizing which part of the advantage is ensemble-dependent. The ratio of perturbative to conventional phase range is $\sim\varepsilon$ independently of the target ensemble. Only the additional $1/\sqrt{N}$ contraction with size requires a high-entropy target distribution. Specifically, one whose differential entropy lies within an additive $O(N^2)$ constant of the Ginibre value, such as Haar-random unitary matrices.

\subsection{Losses introduced by the perturbative programming}\label{sec:losses}

The perturbative MVM method enables a substantially reduced modulation range of the programmable elements, but this reduction comes at the expense of an intrinsic throughput penalty. In the scheme of Fig.~\ref{fig:2}, the desired matrix is realized only up to an amplitude prefactor $\varepsilon/2<1$, so the useful output field is attenuated even before physical insertion losses of components are considered.

This attenuation can be expressed as an equivalent insertion loss,
\begin{equation}
    \mathrm{IL}^{(\mathrm{pert})}(\varepsilon)=
    -20\log_{10}\left(\frac{\varepsilon}{2}\right)\,[\mathrm{dB}],
    \label{eq:ILpert}
\end{equation}
which increases monotonically as $\varepsilon$ decreases. Since $\varepsilon$ is bounded by Eq.~\eqref{eqn:eps_max}, the smallest possible overhead for the two-unitary construction is
\begin{equation}
    \mathrm{IL}^{(\mathrm{pert})}_{\min}
    = \mathrm{IL}^{(\mathrm{pert})}(\varepsilon_{\max})
    \approx 16.7~\mathrm{dB}.
\end{equation}

This is substantial. Nevertheless, a net gain may still arise in regimes where the dominant physical loss scales with the required phase-tuning range, most notably when propagation loss is dominated by long programmable phase shifters. Let $\delta_{2\pi}$ denote the insertion loss of a phase shifter designed for a full $2\pi$ modulation range. If a particular decomposition instance requires a maximal programmed phase excursion $2x_{\max}(\varepsilon)$, then the loss per phase shifter is modeled as

\begin{equation}
    \mathrm{IL}_{\mathrm{ps}}(\varepsilon)
    =
    \delta_{2\pi}\frac{x_{\max}(\varepsilon)}{\pi}.
    \label{eq:ILps}
\end{equation}

\begin{figure}[t]
    \centering
    \includegraphics[width=\linewidth]{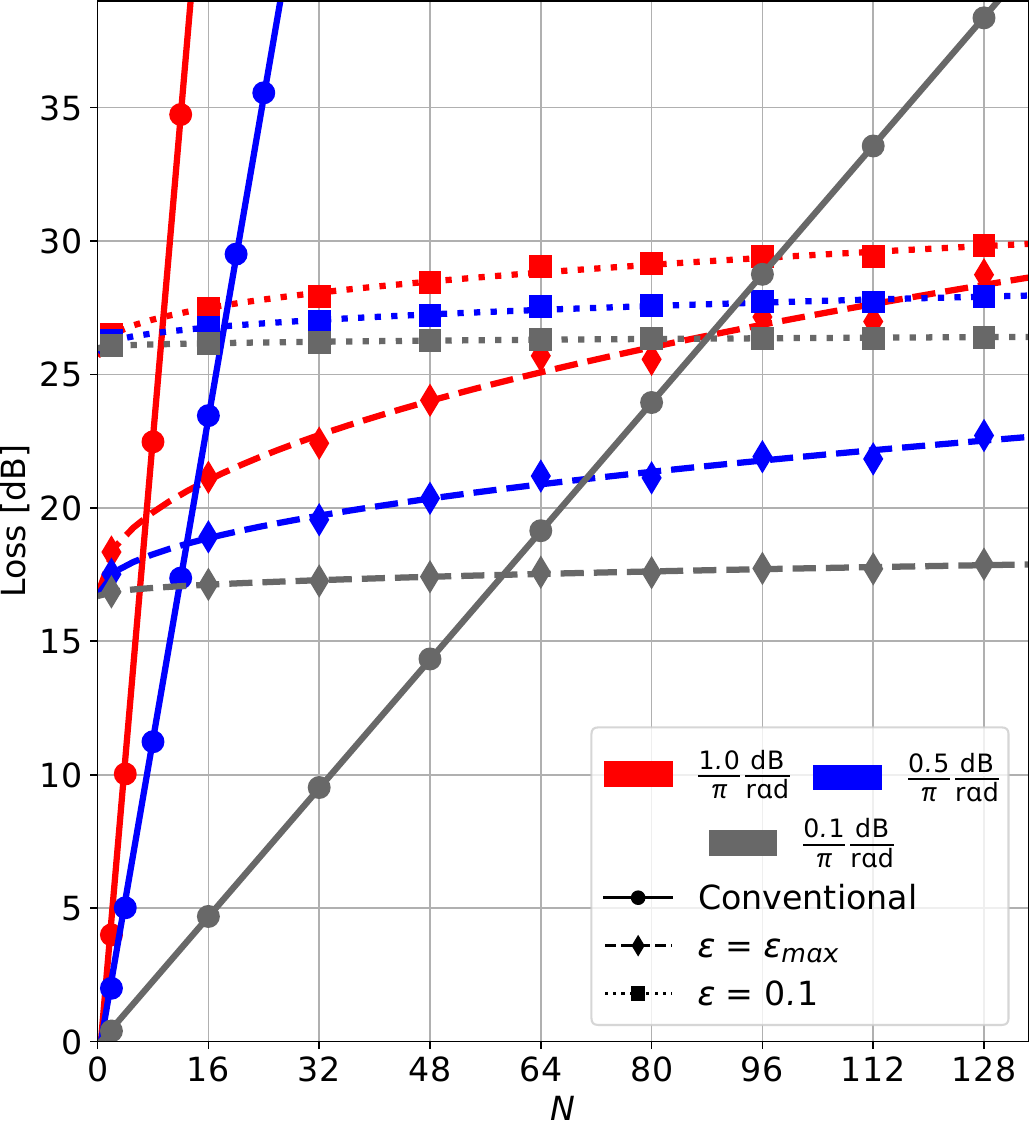}
    \caption{\textbf{Losses introduced by the all-optical perturbative MVM method}: Comparison of per-channel losses for a conventional two-unitary decomposition and perturbative decompositions at different levels of $\varepsilon$ and $\delta_{2\pi}$. The target matrices were sampled from the Ginibre ensemble and then normalized.}
    \label{fig:6}
\end{figure}

In practice, the per-channel programmable loss is obtained by summing Eq.~\eqref{eq:ILps} over all tunable phase shifters encountered along the optical path; this total scales linearly with circuit depth and therefore increases with matrix dimension. The perturbative method is thus governed by a competition between a fixed subtraction overhead and a dimension-dependent reduction in programmable loss.

Fig.~\ref{fig:6} quantifies this balance by comparing the per-channel loss of a conventional non-perturbative two-unitary realization against perturbative realizations for different values of $\varepsilon$. We show three representative phase shifter loss parameters, $\delta_{2\pi}\in\{0.1,0.5,1.0\}$ dB, spanning aggressively optimized low-loss platforms, state-of-the-art thin-film lithium-niobate modulators, and more lossy mass-manufacturable platforms such as silicon photonics. For small matrix dimensions, the intrinsic perturbative overhead dominates and the perturbative method provides no net benefit. As the dimension increases, the savings due to shorter required phase shifters accumulate and can eventually compensate the fixed overhead. The crossover occurs at smaller $N$ when $\delta_{2\pi}$ is larger. While Fig.~\ref{fig:6} uses the Ginibre ensemble of target matrices, the same behavior occurs for Haar random unitaries, with only a shift in the crossover dimensions (verified numerically; not shown).

The subtraction overhead is not strictly immutable. As shown in Appendix~\ref{app:loss}, the balanced 50:50 splitters can be replaced by optimized asymmetric couplers, improving the effective transmission prefactor of the target matrix. This does not alter the basic scaling trade-off, but it can provide a useful constant-factor improvement. Offloading the subtraction to the electronic domain simplifies the implementation: the static part and beamsplitters of Fig.~\ref{fig:2} are no longer needed, and no extra optical power is discarded into dump ports. In this electronic-subtraction variant, the intrincsic overhead of the approach is reduced from the interferometric value $\varepsilon_{\max}/2$ to the signal scaling factor alone $\varepsilon_{\max}$. Importantly, this residual scaling factor is not dissipated optical power, but a reduction of the usefull signal relative to the reference component (inverted input vector).

\section{Discussion and outlook}\label{sec:discussion}

We have proposed a perturbative programming strategy for photonic MVM in programmable interferometer meshes. The central idea is to avoid direct global programming of the target matrix and instead operate the circuit near a fixed reference configuration. The desired transformation is then encoded in the small difference between the programmed transfer matrix and the reference transfer matrix.

Our approach is complementary to existing routes for improving programmable photonic processors. Material and device-level advances reduce the length, voltage, power, or loss required to produce a given phase shift, while phase-efficient mesh architectures reduce the typical phase excursions required by direct programming. The perturbative method adds another layer of optimization: for a fixed programmable architecture, it further reduces the phase range by implementing only a controlled deviation from a reference transformation.

For dense random matrices, our  results show that perturbative programming substantially narrows the phase distributions relative to conventional direct programming. The effect is especially pronounced for the 3-MZI architecture: in the conventional case, the bulk of the phase distribution narrows with matrix size, but rare large phase excursions remain. By contrast, in the perturbative regime the whole distribution, including its near-extreme tails, contracts with increasing $N$. This distinction is important for hardware design, because the physical phase shifter range is usually dictated not only by typical values, but by the largest excursions that must be supported with high yield across the full mesh.

As discussed in Sec.~\ref{subsec:pert mvm}, the asymptotic $1/\sqrt{N}$ contraction is specific to high-entropy ensembles. For such high-entropy target ensembles, the perturbative approach gives the $1/\sqrt{N}$ phase-range scaling discussed in Appendix~\ref{app:sec:entropic}. Sparse and highly structured matrices need not exhibit the same behavior. In particular, the permutation matrix ensemble considered in Appendix~\ref{appendix:sparse}, relevant to optical switching and interconnects, does not exhibit an asymptotic $1/\sqrt{N}$ reduction of the required phase range. In practice, however, little is lost: switching workloads are already well served by dedicated architectures, while accepting a restricted matrix class in exchange for hardware efficiency is itself an established design philosophy, e.g., dense yet non-universal butterfly (FFT) meshes have successfully trained and deployed optical neural networks~\cite{guHardwareEfficientOpticalNeural2021,fengCompactButterflyStyleSilicon2022}. The perturbative approach is less restrictive still: the matrix class remains universal, with the phase-range advantage lost for low-entropy, highly structured ensembles.

The method is not unconditionally advantageous. In the all-optical subtraction scheme, the useful signal carries an intrinsic attenuation factor proportional to $\varepsilon/2$, producing a fixed loss overhead. This penalty is severe for small matrices or platforms with very low phase shifter loss. However, when phase shifter loss scales appreciably with modulation range, the reduced phase excursion can compensate for the subtraction overhead at sufficiently large matrix sizes. The practical benefit is therefore set by the interplay between the perturbation scale $\varepsilon$, the loss per $\pi$ phase shift, and the mesh depth.

A further practical constraint is static hardware error, as discussed in Appendix~\ref{appendix:hardware}. Imperfect directional couplers and fabrication-induced phase offsets can shift the calibrated reference point. In 3-MZI blocks, these errors can be compensated to leading order by trimming the internal phases, preserving the asymptotic phase-range reduction for dense matrices. However, the phase shifters must still provide enough range to reach the corrected operating point before the perturbative modulation is applied. Thus, the achievable reduction is ultimately limited by fabrication accuracy, trimming precision, and long-term drift.

In summary, perturbative programming provides a route to reducing the programmable phase excursion required for photonic MVM. For the two-unitary architecture studied here, it narrows the phase distributions by an amount controlled by the perturbation scale $\varepsilon$ and, when combined with 3-MZI meshes, leads to contraction of the full phase distribution with increasing matrix size. This reduction can translate into lower phase-shifter loss and shorter programmable elements, but only when it outweighs the subtraction overhead and remains above the residual hardware-error floor. The method therefore identifies a new design trade-off for scalable programmable photonics: replacing large-range global programmability by calibrated small-signal programmability around a carefully chosen reference transformation. Although we have focused on integrated interferometer meshes, the perturbative principle is not specific to this platform. It can be applied more generally to any programmable optical system in which a transfer matrix can be biased near a reference configuration and small deviations from that reference can be controlled, including non-planar integrated architectures and free-space optical processors.

\begin{acknowledgments}
S.A.F. acknowledges the support from the Foundation for the Advancement of Theoretical Physics and Mathematics (BASIS) (Project № 23-2-10-15-1) and the Scholarships of the President of the Russian Federation for postgraduate students.
\end{acknowledgments}

\appendix

\section{Deriving optimal $A^{(0)}$}\label{app:A0}

In these derivations we make extensive use of the Haar measure~\cite{Diaconis_2017,Isaev_2018}. The explicit form of the Haar measure in the Clements MZI parameterization of unitary matrices has been identified previously in Refs.~\cite{Russell_2017,Pai_2019}.

\subsection{Optimal unitary}

We begin with the differential of a unitary matrix, $dU$. By vectorizing $dU$, i.e., concatenating its columns into a complex vector $\mathrm{vec}[dU]$ of size $N^2$, we obtain
\begin{equation}
    \mathrm{vec}[dU]=
    \frac{\partial\,\mathrm{vec}[U(\boldsymbol{x})]}{\partial \boldsymbol{x}}d\boldsymbol{x},
\end{equation}
where $\boldsymbol{x}$ denotes a real parameterization of the unitary matrix. To maximize the scope of realizable differentials $dU$ while minimizing $d\boldsymbol{x}$, we seek to maximize the Jacobian magnitude
\begin{equation}
    \left|
    \frac{\partial\,\mathrm{vec}[U(\boldsymbol{x})]}{\partial \boldsymbol{x}}
    \right|.
\end{equation}
This expression is proportional to the Haar measure of the unitary group.

Since the Haar measure is unique up to a positive multiplicative constant \cite{v.neumanUniquenessHaarsMeasure1936, Isaev_2018}, we may directly use the explicit expressions derived in \cite{Russell_2017, Pai_2019} for the Clements MZI parametrization of
\begin{equation}
    \left| \frac{\partial \mathrm{vec}[U]}{\partial (\boldsymbol{\varphi},\boldsymbol{\theta},\boldsymbol{\gamma})}\right|.
\end{equation}
The corresponding measure is
\begin{equation}
    \prod_{i=1}^{\frac{N(N-1)}{2}}\!\!\!\!\!\alpha_i\sin\frac{\theta_i}{2}\left(\cos\frac{\theta_i}{2}\right)^{2\alpha_i-1}d\theta_i \frac{d\varphi_i}{2\pi}\prod_{j=1}^{N}\frac{d\gamma_j}{2\pi},
\end{equation}
where $\alpha_i$ is the location-dependent integer ``sensitivity index'' \cite{Pai_2019}, and $\gamma_j$ denotes the diagonal phase shifts at the inputs/outputs of the Clements decomposition. Maximizing this expression shows that $\varphi_i$ and $\gamma_j$ are uniformly distributed, while
\begin{equation}
    \left(\sin\frac{\theta_i}{2}\right)^2 = \frac{1}{2\alpha_i}.
\end{equation}

To determine the optimal phases for the 3-MZI parametrization, we perform a change of variables, following the same general procedure as in \cite{HamerlyLimitPhases, Hamerly3MZI}. This leads to the maximization of
\begin{equation}
    \prod_{i=1}^{\frac{N(N-1)}{2}}\!\!\!\!\!\frac{\alpha_i\sin{\theta_i}\left(1-\sin\theta_i\sin\varphi_i\right)^{\alpha_i-1}}{2^{\alpha_i+1}\pi}\,d\theta_i\,d\varphi_i\prod_{j=1}^{N}\frac{d\gamma_j}{2\pi},
\end{equation}
which attains its maximum at
\begin{equation}
    \theta_i^{(0)} = -\varphi_i^{(0)} = \frac{\pi}{2},
\end{equation}
which is equivalent to setting a constant phase-shift term, consistent  with the 3-MZI configuration shown in Fig.~\ref{fig:3}, while the phases $\gamma_j$ remain free parameters. Accordingly, at the optimum, each 3-MZI functions as a crossing, such that $U^{(0)}$ becomes an antidiagonal permutation.

\subsection{Optimal pair of unitaries}\label{app:pair_unitary}

We now derive the condition on the reference matrix $A^{(0)}$ for the two-unitary design; see Fig.~\ref{fig:1}c. Consider the differential of
\begin{equation}
    A = \frac{U_1 + U_2}{2}.
\end{equation}
Realizing an arbitrary infinitesimal matrix $dA \sim W$ is equivalent to realizing an arbitrary infinitesimal matrix $dB \sim U_1^\dagger W$, namely
\begin{equation}
    2dB = U_1^\dagger dU_1 + U_1^\dagger U_2 U_2^\dagger dU_2.
\end{equation}
Since unitary matrices satisfy $U_i^\dagger U_i = I$, it follows that $U_i^\dagger dU_i$ is an arbitrary infinitesimal anti-Hermitian matrix. Defining
\begin{equation}
    dG_i = U_i^\dagger dU_i,
    \qquad
    Z = U_1^\dagger U_2,
\end{equation}
we obtain
\begin{equation}
    2dB = dG_1 + Z dG_2.
\end{equation}

On the left-hand side, $dB$ is a matrix of independent differentials, which allows its Hermitian and anti-Hermitian parts to be treated independently. In particular,
\begin{equation}
    2(dB + dB^\dagger) = Z dG_2 - dG_2 Z^\dagger.
\end{equation}
After vectorization, the right-hand side becomes
\begin{equation}
    (I \otimes Z - Z^* \otimes I)\,\mathrm{vec}[dG_2].
\end{equation}
To optimize the operating point, we therefore choose $Z$ so as to maximize the determinant magnitude
\begin{equation}
    \left|I \otimes Z - Z^* \otimes I\right|.
\end{equation}

An upper bound on the determinant of a sum of matrices can be expressed in terms of the singular values of the summands \cite{Li_1995}. Since $Z$ is unitary, all of its singular values are equal to $1$, which gives
\begin{equation}
    \left|I \otimes Z - Z^* \otimes I\right|
    \leq
    \prod_{i=1}^{N^2}(1+1)
    =
    2^{N^2}.
\end{equation}
This bound is saturated when
\begin{equation}
    Z = \pm iI,
\end{equation}
which implies
\begin{equation}
    U_2 = \pm i U_1.
\end{equation}
Thus, this choice provides the optimal operating point. Taking the plus sign, the optimal reference matrix can be written as
\begin{equation}\label{app:eq:A0}
    A^{(0)}=\frac{e^{i\pi/4}}{\sqrt{2}}\,U^{(0)},
\end{equation}
where $U^{(0)}$ is a unitary antidiagonal permutation matrix.

\section{Phase distributions for the conventional and perturbative regimes }\label{appendix:relations}

Approximating the inequalities in Eqs.~\eqref{app:eq:entropy subadd marginal} and \eqref{app:eq:moment lower bound} as equalities yields the relation:
\begin{equation}
    \frac{\langle x^{(pert)}  \rangle_p}{\langle x^{(conv)} \rangle_p}\sim e^{\Delta H_{\boldsymbol{x}}/2N^2}
\end{equation}
In the conventional two-unitary architecture, the differential
\begin{equation}
    dA = \frac{dU_1+dU_2}{2}
\end{equation}
is embedded into a larger differential of unitary $d\mathcal{U}$ given by:
\begin{equation}\label{app:eq:embedding two-unitary}
    d\mathcal{U} = \frac{1}{2}
    \begin{pmatrix}
        dU^{(1)}+dU^{(2)} & dU^{(1)}-dU^{(2)} \\
        dU^{(1)}-dU^{(2)} & -\left(dU^{(1)}+dU^{(2)}\right) 
    \end{pmatrix}.
\end{equation}
Following the same reasoning as in Appendix~\ref{app:sec:entropic}, we obtain
\begin{equation}
    2\|dA\|=\|d\mathcal{U}\|
\end{equation}
and
\begin{equation}\label{app:eq:jac conv}
    \left|\frac{\partial \mathrm{vec}\left[A\right] }{\partial \boldsymbol{x}}\right|\leq\left(\frac{1}{2}\right)^{2N^2}.
\end{equation}
Comparing this result with Eq.~\eqref{app:eq:jac pert} yields:
\begin{equation}
    \frac{\langle x^{(pert)}  \rangle_p}{\langle x^{(conv)} \rangle_p}\sim\varepsilon,
\end{equation}
indicating that the perturbative bound is approximately a fraction $\varepsilon$ of the conventional bound. Provided that approximation of a true value with a bound is true.

\section{Entropic bound}\label{app:sec:entropic}

Here we follow the framework developed in \cite{HamerlyLimitPhases}. Let $W$ be a target $N\times N$ complex matrix, and let $\boldsymbol{x}=(\boldsymbol{x}_1,\boldsymbol{x}_2)$ denote the $2N^2$ parameters of its two-unitary decomposition. We estimate the entropy as
\begin{equation}\label{app:eq:Hpsi}
    H_{\boldsymbol{x}}=H_W-\left\langle \log \left|\frac{\partial \mathrm{vec}[W]}{\partial \boldsymbol{x}}\right| \right\rangle,
\end{equation}
where vectorization is understood to concatenate both the real and imaginary parts.

To assess the fabrication requirements for the phase shifters, we consider the compound distribution of all phase variables,
\begin{equation}
    \rho(x)=\frac{1}{2N^2}\sum_{i=1}^{2N^2}\rho_i(x),
\end{equation}
where $\rho_i(x)$ is the marginal distribution of the phase at site $i$. Using the concavity and subadditivity of entropy, we obtain
\begin{equation}\label{app:eq:entropy subadd marginal}
    H(\rho)\geq \frac{1}{2N^2}\sum_{i=1}^{2N^2}H(\rho_i)\geq \frac{H_{\boldsymbol{x}}}{2N^2}.
\end{equation}
Applying the maximum entropy principle, subject to the constraint on the $p$-norm $\langle x \rangle_p=\left(\int dx |x|^p \rho(x)\right)^{1/p}$ of the compound distribution, yields an upper bound on the entropy $H(\rho)$ \cite{Jaynes1957, HamerlyLimitPhases}:
\begin{equation}
    H(\rho)\leq \ln\left(2\Gamma(1/p+1)e^{1/p} p^{1/p} \langle x \rangle_p\right),
\end{equation}
resulting in a lower bound on $\langle x\rangle_p$
\begin{equation}\label{app:eq:moment lower bound}
    \langle x \rangle_p\geq \frac{e^{H(\rho)-1/p}}{2p^{1/p}\Gamma(1/p+1)}.
\end{equation}
This result consequently establishes a lower bound for
the maximal phase element ($p=\infty$):
\begin{equation}\label{app:eq:max bound entropy}
    \max |x|\geq \frac{1}{2}e^{H_{\boldsymbol{x}}/2N^2},
\end{equation}
which defines the parameter range $[-\max|x|,\max|x|]$. To reconstruct $H_{\boldsymbol{x}}$ from Eq.~\eqref{app:eq:Hpsi}, we therefore need both $H_W$ and the average Jacobian.

Our target matrices $W$ can be obtained from Ginibre-ensemble matrices $G$ satisfying
\begin{equation}
    \langle G_{ij}\rangle=0,
    \qquad
    \langle \Re[G_{ij}]^2\rangle
    =
    \langle \Im[G_{ij}]^2\rangle
    =
    \frac{1}{8N},
\end{equation}
and then normalized by their largest singular value,
\begin{equation}
    W=\frac{G}{\|G\|_2}.
\end{equation}
The variance $1/(8N)$ is chosen so that, for $N\gg 1$, the Mar\v{c}enko--Pastur quarter-circle law implies $\|G\|_2\approx 1$ \cite{Marcenko1967,Bai2010}, via the standard connection between the eigenvalues of $GG^\dagger$ and the singular values of $G$.

The entropy of the Ginibre ensemble is simply the sum of the entropies of the Gaussian-distributed matrix elements,
\begin{equation}
    H_G=N^2\log\left(\frac{\pi e}{4N}\right).
\end{equation}
What we actually need, however, is $H_W$. On heuristic grounds, we argue that the difference $H_G-H_W$ grows more slowly than $O(N^2)$, so that the leading order approximation
\begin{equation}
    H_W\approx H_G
\end{equation}
is sufficient for our purposes.

To estimate the Jacobian, we recall the perturbative decomposition
\begin{equation}\label{app:eq:perturb}
    \varepsilon W = \frac{(U_1-U^{(0)}_1)+(U_2-U^{(0)}_2)}{2}.
\end{equation}
Multiplying $(U_1-U^{(0)}_1)+(U_2-U^{(0)}_2)$ on the left by the constant unitary matrix $U^{(0)\dagger}_1$, we obtain
\begin{multline}
    U^{(0)\dagger}_1\bigl((U_1-U^{(0)}_1)+(U_2-U^{(0)}_2)\bigr)
    =\\
    U^{(0)\dagger}_1(U_1-U^{(0)}_1)+i\,U^{(0)\dagger}_2(U_2-U^{(0)}_2),
\end{multline}
where we used the relation $U^{(0)}_2=iU^{(0)}_1$, derived in Appendix~\ref{app:pair_unitary}. To first order in $\varepsilon$, the terms $U^{(0)\dagger}_i(U_i-U^{(0)}_i)$ belong to the Lie algebra $\mathfrak{su}(N)$ \cite{Isaev_2018} and are therefore anti-Hermitian. Consequently, the same is true for their differentials $U^{(0)\dagger}_i dU_i$.

As shown in Fig.~\ref{fig:2}, the perturbative two-unitary construction embeds the target matrix $\frac{\varepsilon}{2}W$ into a larger matrix $\mathcal{U}$. The corresponding differential is
\begin{equation}\label{app:eq:embedding}
    d\mathcal{U} = \frac{1}{2}
    S\begin{pmatrix}
    dU^{(1)}+dU^{(2)} & dU^{(1)}-dU^{(2)} & 0\\
    dU^{(1)}-dU^{(2)} & -\left(dU^{(1)}+dU^{(2)}\right) & 0\\
    0 & 0 & 0
    \end{pmatrix}S,
    \end{equation}
with unitary $S$ being outermost splitters' matrices. 

Since the squared Frobenius norm of a sum of Hermitian and anti-Hermitian matrices splits additively, we have
\begin{multline}
    \|dU_1\pm dU_2\|^2
    =
    \left\|U^{(0)\dagger}_1(dU_1\pm dU_2)\right\|^2
    =\\
    \|U^{(0)\dagger}_1 dU_1\|^2 + \|U^{(0)\dagger}_2 dU_2\|^2
    =
    \|dU_1\|^2+\|dU_2\|^2.
\end{multline}
Equation~\eqref{app:eq:perturb} therefore implies
\begin{equation}
    \|dU_1\pm dU_2\| = 2\varepsilon \|dW\|.
\end{equation}
Substituting this into Eq.~\eqref{app:eq:embedding}, we obtain
\begin{equation}
    \|d\mathcal{U}\| = 2\varepsilon \|dW\|.
\end{equation}

Phase shifters are represented by diagonal matrices with entries $e^{ix}$, sandwiched between unitary matrices. Consequently, varying a single phase shifter gives
\begin{equation}
    \|dx\| = \|d\mathcal{U}\|,
\end{equation}
which leads to the Jacobian bound
\begin{equation}\label{app:eq:jac pert}
    \left|\frac{\partial \mathrm{vec}[W]}{\partial \boldsymbol{x}}\right|
    \leq
    \left(\frac{1}{2\varepsilon}\right)^{2N^2}.
\end{equation}
As a result, the entropy $H_{\boldsymbol{x}}$ satisfies
\begin{multline}
    H_{\boldsymbol{x}}
    \geq
    N^2\log\left(\frac{\pi e}{4N}\right)+2N^2\log(2\varepsilon)
    =\\
    2N^2\log\left(\varepsilon\sqrt{\frac{\pi e}{N}}\right).
\end{multline}
Finally, for target matrices sampled from the Ginibre ensemble and normalized by their spectral norm, we obtain
\begin{equation}\label{app:eq:entropic_bound}
    \max|x| \geq \varepsilon \sqrt{\frac{\pi e}{4 N}}.
\end{equation}
Numerically, we find
\begin{equation}
    \max|x|\bigg/\left(\varepsilon \sqrt{\frac{\pi e}{4N}}\right)\sim 3.8.
\end{equation}

\section{Sparse-matrix phase distribution}\label{appendix:sparse}

Substituting Eqs.~\eqref{app:eq:Hpsi} and \eqref{app:eq:jac pert} into Eq.~\eqref{app:eq:max bound entropy}, we obtain the general lower bound
\begin{equation}
    \max |x| \geq \varepsilon\, e^{H_W/2N^2}.
    \label{app:eq:sparse_general_bound}
\end{equation}
This expression shows that the entropy of the target-matrix ensemble determines whether the required phase range decreases asymptotically with matrix size. For dense Ginibre matrices, the entropy scales as $H_W\simeq N^2\log(\pi e/4N)$, which leads to the $1/\sqrt{N}$ behavior derived in Eq.~\eqref{app:eq:entropic_bound}. This scaling, however, is not universal and depends on the distribution from which the target matrices are drawn.

As a counterexample, consider the ensemble of permutation matrices $P$. In the perturbative 3-MZI architecture, the Jacobian matrix and its inverse exhibit significant sparsity at the optimal operating point $A^{(0)} = A(\boldsymbol{x}^{(0)})$. For sufficiently small $\varepsilon$, the parameters are given by:
\begin{equation}
\Delta \boldsymbol{x} \approx \varepsilon \left[ \left. \frac{\partial \operatorname{vec}\left[A(\boldsymbol{x})\right]}{\partial \boldsymbol{x}} \right|_{\boldsymbol{x}^{(0)}} \right]^{-1} \cdot \operatorname{vec}\left[P\right].
\end{equation}
The inverse Jacobian has a specific structure: the rows contain at most two non-zero entries, with modulo equal to $2$. Since $\operatorname{vec}\left[P\right]$ consists entirely of zeros and ones, each phase shifter requires support for these distinct states. Thus at least:
\begin{equation}
    \max |x| \geq 2\varepsilon,
\end{equation}
up to subleading corrections. Thus, unlike the dense Ginibre case, the permutation matrix ensemble does not exhibit an asymptotic $1/\sqrt{N}$ reduction of the required phase range. This illustrates that the phase-efficiency scaling of the perturbative approach is ensemble-dependent: dense random matrices benefit from the high entropy of the target distribution, whereas highly structured sparse ensembles, such as permutation matrices, need not display the same scaling.

\section{Reducing losses with asymmetric couplers}\label{app:loss}

The scheme of Fig.~\ref{fig:2} can be improved slightly. Since
\begin{equation}
    A^{(0)}=\frac{e^{i\pi/4}}{\sqrt{2}}U^{(0)},
\end{equation}
directly implementing the static portion with the prefactor $1/\sqrt{2}$ would introduce unnecessary attenuation. Instead, one may implement the static transformation $e^{i\pi/4}U^{(0)}$ directly and absorb the factor into modified splitters used for matrix subtraction. Replacing standard 50:50 splitters with asymmetric splitters of ratios
\begin{equation}
    \frac{\sqrt{2}}{1+\sqrt{2}} : \frac{1}{1+\sqrt{2}},
    \qquad
    \frac{1}{1+\sqrt{2}} : \frac{\sqrt{2}}{1+\sqrt{2}},
\end{equation}
yields a prefactor $\varepsilon(2-\sqrt{2})$ instead of $\varepsilon/2$. This corresponds to an increase in signal intensity of approximately $1.38$ dB, at the cost of requiring a less convenient splitting ratio of roughly $59:41$.

\section{Static hardware errors}\label{appendix:hardware}

The preceding analysis assumes ideal interferometric blocks. In practice, however, the constituent directional couplers are not perfectly balanced. Such static imperfections can be modeled, for example, by writing the beamsplitter transformation as
\begin{equation}\label{eqn:err_block}
    U_{\mathrm{BS}}=
    \begin{pmatrix}
        \cos\left(\frac{\pi}{4}+\alpha\right) &
        i\sin\left(\frac{\pi}{4}+\alpha\right)\\
        i\sin\left(\frac{\pi}{4}+\alpha\right) &
        \cos\left(\frac{\pi}{4}+\alpha\right)
    \end{pmatrix},
\end{equation}
where $\alpha$ quantifies the deviation from the ideal $50{:}50$ splitting ratio.

\begin{figure}[hbt!]
	\centering
	\includegraphics[width=0.7\linewidth]{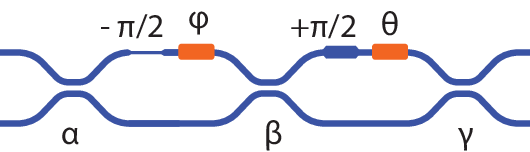}
	\caption{\textbf{Imperfect 3-MZI block.} A 3-MZI with unbalanced directional couplers, characterized by error angles $\alpha$, $\beta$, and $\gamma$ according to Eq.~\eqref{eqn:err_block}. Static coupler errors shift the phase settings required to realize the optimal crossing configuration and must therefore be compensated by post-fabrication trimming.}
	\label{app:fig:hardware error}
\end{figure}

For an imperfect 3-MZI block, shown schematically in Fig.~\ref{app:fig:hardware error}, such beamsplitter errors shift the phase settings required to realize the optimal crossing configuration. In the small-error limit, the necessary compensation phases are
\begin{equation}\label{eqn:err_3mzi_corrections}
	\Delta\varphi = 2\gamma, \qquad
	\Delta\theta = 2\alpha,
\end{equation}
where $\alpha$ and $\gamma$ are the imbalance parameters of the corresponding directional couplers, defined as in Eq.~\eqref{eqn:err_block}. Thus, if beamsplitter or static phase errors are characterized after fabrication, the 3-MZI can be trimmed back to its desired operating point by applying the additional phase shifts in Eq.~\eqref{eqn:err_3mzi_corrections}.

Importantly, this trimming correction is static and slow: it need not be supplied by the fast programmable phase shifter at all. It can instead be absorbed by one-time non-volatile post-fabrication trimming, e.g., via laser or electrical annealing of germanium-implanted waveguide sections~\cite{chenPostfabricationPhaseTrimming2017, loganHighResolutionTrimming2026}, or by a separate low-speed bias section. Quantitatively, the perturbative excursion $x_{\max} \approx 3.8\varepsilon_{\mathrm{max}}\sqrt{\pi e/4N}\approx0.2$ rad at $N = 64$ exceeds current residual trimming errors of $\approx2$ mrad~\cite{loganHighResolutionTrimming2026} by two orders of magnitude. With the static correction offloaded in this way, the fast, low-loss programmable element needs to cover only the perturbative excursion, and the range reduction analyzed above is preserved in full.

This observation is important for the perturbative approach. After such trimming, the asymptotic $1/\sqrt{N}$ scaling of the required phase range for dense matrices is preserved. However, if no dedicated trimming mechanism is available and the correction must be supplied by the programmable phase shifters themselves, the phase range cannot be reduced arbitrarily. Even though the perturbative method decreases the programmable excursion needed around the operating point, the operating point itself must first be reached by compensating fabrication-induced errors. Consequently, the available phase shifter range must then be large enough to cover both the intended perturbative modulation and the static correction required after fabrication.

Therefore, the practically achievable reduction of the phase shifter range is ultimately bounded by fabrication accuracy, trimming precision, and long-term drift. Manufacturing errors in the phase shifters themselves lead to the same conclusion: any uncompensated static phase offset directly increases the minimum required tuning range.

\subsubsection*{Computation accuracy under stochastic phase errors}\label{app:accuracy}

We introduce random static deviations \mbox{$\delta \varphi,\delta\theta \sim \mathcal{N}(0,\sigma^2)$}, independent of the tuning range, to the operating point phases and for erroneous $\boldsymbol{x}$ quantify the normalized MVM error 
\begin{equation}
	\mathcal{E} =\frac{\|(A(\tilde{\boldsymbol{x}})-A_0) - \varepsilon W\|_F}{\varepsilon \|W\|_F}
\end{equation}
This models, e.g., residual fabrication-induced phase offsets remaining after imperfect trimming. The perturbative element compensates the displacement within its reachable range $[-x_{\max}, x_{\max}]$, realizing the closest attainable phase to the ideal value. For comparison, the same deviations are applied to the conventional phases, which are left uncorrected, and $\mathcal{E} = \|\tilde{A} - A\|_F / \|A\|_F$ is extracted.

\begin{figure}[hbt!]
\centering
\includegraphics[width=\linewidth]{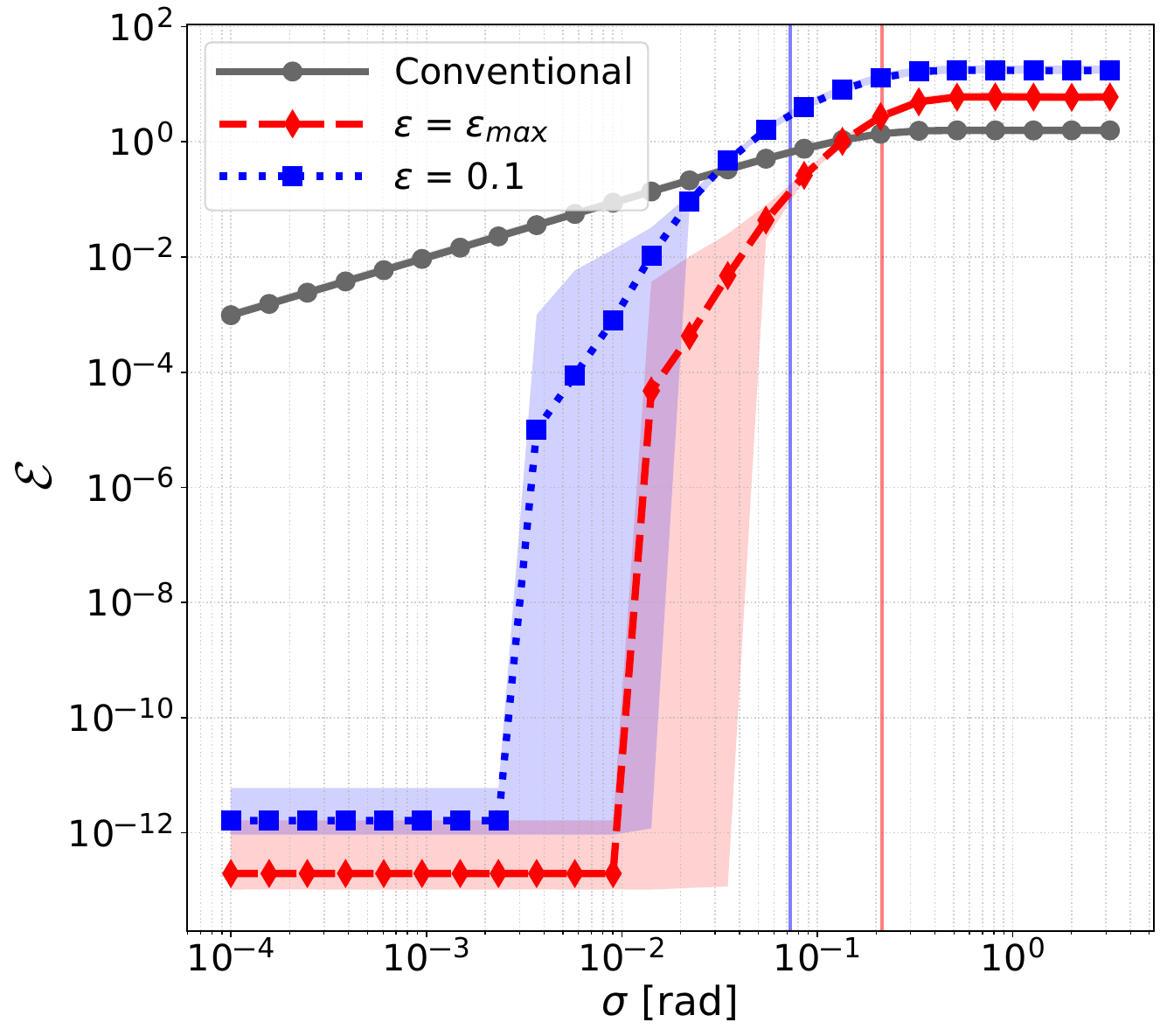}
\caption{\textbf{Computation accuracy under stochastic phase errors.} Normalized MVM error $\mathcal{E}$ versus rms phase error $\sigma$ for the conventional and perturbative 3-MZI schemes ($N=64$, $100$ Ginibre targets), with identical range-independent errors $\delta x_i \sim \mathcal{N}(0,\sigma^2)$. The conventional phases are left uncorrected, whereas the perturbative ones are compensated within the reachable range $[-x_{\max}, x_{\max}]$ ($x_{\max}$ indicated by vertical lines). Lines are means; shaded regions span the full statistics.}
\label{app:fig:stochastic error}
\end{figure}

As shown in Fig.~\ref{app:fig:stochastic error}, once the required correction reaches about a fraction of $x_{\max}$, the residual deviation can no longer be absorbed. The perturbative error is then amplified by a factor $\sim 1/\varepsilon$ relative to the conventional one, because it is referenced to a signal of magnitude $\varepsilon$. Accuracy in the perturbative regime is therefore bounded by the same range floor that governs the phase-range reduction itself. The plateau for small $\sigma$ simply indicates that the random error had the opposite sign to the phase on the site with the maximum phase excursion.

Errors that scale with the programmed swing---DAC quantization at fixed bit depth, drive-voltage noise, crosstalk---instead contract together with $x_{\max}$. At equal control resources, the perturbative scheme is therefore at least as accurate as the conventional one.

\bibliography{biblio.bib}

@article{fengCompactButterflyStyleSilicon2022,
	title = {A {{Compact Butterfly-Style Silicon Photonic}}--{{Electronic Neural Chip}} for {{Hardware-Efficient Deep Learning}}},
	author = {Feng, Chenghao and Gu, Jiaqi and Zhu, Hanqing and Ying, Zhoufeng and Zhao, Zheng and Pan, David Z. and Chen, Ray T.},
	year = 2022,
	month = dec,
	journal = {ACS Photonics},
	volume = {9},
	number = {12},
	pages = {3906--3916},
	issn = {2330-4022, 2330-4022},
	doi = {10.1021/acsphotonics.2c01188},
	url = {https://pubs.acs.org/doi/10.1021/acsphotonics.2c01188},
	urldate = {2026-07-22},
	copyright = {https://doi.org/10.15223/policy-029}
}

@article{guHardwareEfficientOpticalNeural2021,
	title = {Toward {{Hardware-Efficient Optical Neural Networks}}: {{Beyond FFT Architecture}} via {{Joint Learnability}}},
	shorttitle = {Toward {{Hardware-Efficient Optical Neural Networks}}},
	author = {Gu, Jiaqi and Zhao, Zheng and Feng, Chenghao and Ying, Zhoufeng and Liu, Mingjie and Chen, Ray T. and Pan, David Z.},
	year = 2021,
	month = sep,
	journal = {IEEE Transactions on Computer-Aided Design of Integrated Circuits and Systems},
	volume = {40},
	number = {9},
	pages = {1796--1809},
	issn = {0278-0070, 1937-4151},
	doi = {10.1109/TCAD.2020.3027649},
	url = {https://ieeexplore.ieee.org/document/9209122/},
	urldate = {2026-07-22},
	copyright = {https://ieeexplore.ieee.org/Xplorehelp/downloads/license-information/IEEE.html}
}

@article{chenPostfabricationPhaseTrimming2017,
	title = {Post-Fabrication Phase Trimming of {{Mach}}--{{Zehnder}} Interferometers by Laser Annealing of Germanium Implanted Waveguides},
	author = {Chen, Xia and Milosevic, Milan M. and Thomson, David J. and Khokhar, Ali Z. and Franz, Yohann and Runge, Antoine F. J. and Mailis, Sakellaris and Peacock, Anna C. and Reed, Graham T.},
	year = 2017,
	month = dec,
	journal = {Photonics Research},
	volume = {5},
	number = {6},
	pages = {578},
	issn = {2327-9125},
	doi = {10.1364/PRJ.5.000578},
	url = {https://opg.optica.org/abstract.cfm?URI=prj-5-6-578},
	urldate = {2026-07-22}
}

@article{loganHighResolutionTrimming2026,
	title = {Towards {{High Resolution Trimming}} of {{Silicon Photonic Waveguides}}},
	author = {Logan, April M. and Chen, Xia and Yu, Xingshi and Shen, Weihong and Yan, Xingzhao and Zhang, Weiwei and Mashanovich, Goran Z. and Reed, Graham T. and Thomson, David J.},
	year = 2026,
	month = feb,
	journal = {Advanced Photonics Research},
	volume = {7},
	number = {2},
	pages = {e202500298},
	issn = {2699-9293, 2699-9293},
	doi = {10.1002/adpr.202500298},
	url = {https://advanced.onlinelibrary.wiley.com/doi/10.1002/adpr.202500298},
	urldate = {2026-07-22},
	copyright = {http://creativecommons.org/licenses/by/4.0/}
}

@article{BTO_RR,
    author = {Raju, Amogh and Hungund, Divya and Krueger, Dan and Dong, Zuoming and Sakotic, Zarko and Posadas, Agham B. and Demkov, Alexander A. and Wasserman, Daniel},
    title = {{High-{Q} Monolithic Ring Resonators in Low-Loss Barium Titanate on Silicon}},
    journal = {Laser \& Photonics Reviews},
    year = {2025},
    pages = {2402086},
    doi = {https://doi.org/10.1002/lpor.202402086},
    url = {https://onlinelibrary.wiley.com/doi/abs/10.1002/lpor.202402086},
}

@article{Jaynes1957,
  title = {{Information Theory and Statistical Mechanics}},
  author = {Jaynes, E. T.},
  journal = {Physical Review},
  volume = {106},
  number = {4},
  pages = {620--630},
  year = {1957},
  month = {May},
  publisher = {American Physical Society},
  doi = {10.1103/PhysRev.106.620},
  url = {https://link.aps.org/doi/10.1103/PhysRev.106.620}
}

@article{Marcenko1967,
  title = {{Distribution of Eigenvalues for Some Sets of Random Matrices}},
  author = {Mar{\v c}enko, V. A. and Pastur, L. A.},
  year = 1967,
  month = apr,
  journal = {Mathematics of the USSR-Sbornik},
  volume = {1},
  number = {4},
  pages = {457--483},
  issn = {0025-5734},
  doi = {10.1070/SM1967v001n04ABEH001994},
  url = {https://www.mathnet.ru/eng/sm4101},
  urldate = {2026-01-08}
}

@book{Bai2010,
  title = {Spectral Analysis of Large Dimensional Random Matrices},
  ISBN = {9781441906618},
  ISSN = {0172-7397},
  url = {http://dx.doi.org/10.1007/978-1-4419-0661-8},
  DOI = {10.1007/978-1-4419-0661-8},
  series = {Springer Series in Statistics},
  publisher = {Springer New York},
  author = {Bai,  Zhidong and Silverstein,  Jack W.},
  year = {2010}
}

@INPROCEEDINGS{BTO_SI_modulator,
    author={Lievens, E. and Picavet, E. and De Geest, K. and De Buysser, K. and Van Thourhout, D. and Bienstman, P. and Beeckman, J.},
    booktitle={2024 IEEE Photonics Conference (IPC)}, 
    title={{Integration of Barium Titanate Thin Films in Silicon Photonics for Electro-Optic Modulation}}, 
    year={2024},
    pages={1--2},
    doi={10.1109/IPC60965.2024.10799854}
}

@article{BTO_LNOI,
    author = {Wen, Yiyang and Chen, Haisheng and Wu, Zhenping and Li, Wei and Zhang, Yang},
    title = {{Fabrication and photonic applications of {Si}-integrated {LiNbO$_3$} and {BaTiO$_3$} ferroelectric thin films}},
    journal = {APL Materials},
    volume = {12},
    number = {2},
    pages = {020601},
    year = {2024},
    month = {02},
    issn = {2166-532X},
    doi = {10.1063/5.0192018},
    url = {https://doi.org/10.1063/5.0192018},
}

@article{STO_Pockels,
   title={{Engineering high Pockels coefficients in thin-film strontium titanate for cryogenic quantum electro-optic applications}},
   volume={390},
   ISSN={1095-9203},
   url={http://dx.doi.org/10.1126/science.adx3741},
   DOI={10.1126/science.adx3741},
   number={6771},
   journal={Science},
   publisher={American Association for the Advancement of Science (AAAS)},
   author={Ulrich, Anja and Brahim, Kamal and Boelen, Andries and Debaets, Michiel and Khalil, Ahmed and Sun, Conglin and Huang, Yishu and Saseendran, Sandeep Seema and Baryshnikova, Marina and Favia, Paola and Nuytten, Thomas and Sergeant, Stefanie and Van Gasse, Kasper and Kuyken, Bart and De Greve, Kristiaan and Merckling, Clement and Haffner, Christian},
   year={2025},
   month=Oct,
   pages={390--393}
}

@article{Xbar, 
    author={Giamougiannis, George and Tsakyridis, Apostolos and Ma, Yangjin and Totović, Angelina and Moralis-Pegios, Miltiadis and Lazovsky, David and Pleros, Nikos}, 
    journal={Journal of Lightwave Technology},
    title={{A Coherent Photonic Crossbar for Scalable Universal Linear Optics}},
    year={2023}, 
    volume={41}, 
    number={8}, 
    pages={2425-2442},
    doi={10.1109/JLT.2023.3234689}
}

@ARTICLE{WeightBanks,
  author={Tait, Alexander N. and Wu, Allie X. and de Lima, Thomas Ferreira and Zhou, Ellen and Shastri, Bhavin J. and Nahmias, Mitchell A. and Prucnal, Paul R.},
  journal={IEEE Journal of Selected Topics in Quantum Electronics}, 
  title={{Microring Weight Banks}}, 
  year={2016},
  volume={22},
  number={6},
  pages={312-325},
  doi={10.1109/JSTQE.2016.2573583}
}

@article{TwoUnitary,
  title = {{Low-Depth Two-Unitary Design of Programmable Photonic Circuits}},
  author = {Fldzhyan, S. A. and Saygin, M. {\relax Yu} and Straupe, S. S.},
  year = 2026,
  month = jan,
  journal = {Physical Review Research},
  volume = {8},
  number = {1},
  pages = {013021},
  issn = {2643-1564},
  doi = {10.1103/9fg1-fdfl},
  url = {https://link.aps.org/doi/10.1103/9fg1-fdfl},
  urldate = {2026-01-12},
}

@article{ClementsDesign,
  title = {{Optimal Design for Universal Multiport Interferometers}},
  author = {Clements, William R. and Humphreys, Peter C. and Metcalf, Benjamin J. and Kolthammer, W. Steven and Walsmley, Ian A.},
  year = 2016,
  month = dec,
  journal = {Optica},
  volume = {3},
  number = {12},
  pages = {1460},
  issn = {2334--2536},
  doi = {10.1364/OPTICA.3.001460},
  url = {https://opg.optica.org/abstract.cfm?URI=optica-3-12-1460},
  urldate = {2025-12-28},
  copyright = {https://creativecommons.org/licenses/by/4.0/}
}

@Article{Hamerly3MZI,
    author={Hamerly, Ryan
    and Bandyopadhyay, Saumil
    and Englund, Dirk},
    title={{Asymptotically fault-tolerant programmable photonics}},
    journal={Nature Communications},
    year={2022},
    month={Nov},
    day={29},
    volume={13},
    number={1},
    pages={6831},
    issn={2041-1723},
    doi={10.1038/s41467-022-34308-3},
    url={https://doi.org/10.1038/s41467-022-34308-3}
}

@article{HamerlyLimitPhases,
  title = {{Toward the Information-Theoretic Limit of Programmable Photonics}},
  author = {Hamerly, Ryan and Basani, Jasvith R. and Sludds, Alexander and Vadlamani, Sri K. and Englund, Dirk},
  date = {2025-11-01},
  journal = {APL Photonics},
  volume = {10},
  number = {11},
  pages = {110803},
  issn = {2378-0967},
  doi = {10.1063/5.0269741},
  url = {https://pubs.aip.org/app/article/10/11/110803/3370635/Toward-the-information-theoretic-limit-of},
  urldate = {2025-12-28}
}

@article{SVD_Miller,
  title = {{Self-Configuring Universal Linear Optical Component}},
  author = {Miller, David A B},
  date = {2013-06},
  journal = {Photonics Research},
  volume = {1},
  number = {1},
  pages = {1--15},
  doi = {10.1364/PRJ.1.000001}
}

@article{Diaconis_2017,
author = {Diaconis, Persi and Forrester, Peter J.},
title = {{Hurwitz and the origins of random matrix theory in mathematics}},
journal = {Random Matrices: Theory and Applications},
volume = {6},
number = {1},
pages = {1730001},
year = {2017},
doi = {10.1142/S2010326317300017},
URL = {https://doi.org/10.1142/S2010326317300017}
}

@article{Russell_2017,
  title = {{Direct Dialling of {{Haar}} Random Unitary Matrices}},
  author = {Russell, Nicholas J and Chakhmakhchyan, Levon and O'Brien, Jeremy L and Laing, Anthony},
  year = 2017,
  month = mar,
  journal = {New Journal of Physics},
  volume = {19},
  number = {3},
  pages = {033007},
  issn = {1367-2630},
  doi = {10.1088/1367-2630/aa60ed},
  url = {https://iopscience.iop.org/article/10.1088/1367-2630/aa60ed},
  urldate = {2025-12-28}
}

@article{Pai_2019,
  title = {{Matrix Optimization on Universal Unitary Photonic Devices}},
  volume = {11},
  ISSN = {2331-7019},
  url = {http://dx.doi.org/10.1103/PhysRevApplied.11.064044},
  number = {6},
  journal = {Physical Review Applied},
  publisher = {American Physical Society (APS)},
  author = {Pai,  Sunil and Bartlett,  Ben and Solgaard,  Olav and Miller,  David A. B.},
  year = {2019},
  month = jun 
}

@article{Li_1995,
  title = {{The determinant of the sum of two matrices}},
  volume = {52},
  ISSN = {1755-1633},
  url = {http://dx.doi.org/10.1017/S0004972700014908},
  DOI = {10.1017/s0004972700014908},
  number = {3},
  journal = {Bulletin of the Australian Mathematical Society},
  publisher = {Cambridge University Press (CUP)},
  author = {Li, Chi-Kwong and Mathias, Roy},
  year = {1995},
  month = dec,
  pages = {425--429}
}

@article{v.neumanUniquenessHaarsMeasure1936,
  title = {{The Uniqueness of {{Haar}}'s Measure}},
  author = {{v. Neuman}, J.},
  year = 1936,
  journal = {Recueil Math\'ematique (Nouvelle s\'erie)},
  volume = {1(43)},
  number = {5},
  pages = {721--734},
  url = {https://www.mathnet.ru/eng/sm5481}
}

@book{Isaev_2018,
author = {Isaev, Alexey P. and Rubakov, Valery A.},
title = {Theory of Groups and Symmetries},
publisher = {World Scientific},
year = {2018},
doi = {10.1142/10898},
address = {},
edition   = {},
URL = {https://www.worldscientific.com/doi/abs/10.1142/10898},
}

@article{Bogaerts2020,
  author  = {Bogaerts, Wim and P{\'e}rez, Daniel and Capmany, Jos{\'e} and Miller, David A. B. and Poon, Joyce and Englund, Dirk and Morichetti, Francesco and Melloni, Andrea},
  title   = {{Programmable photonic circuits}},
  journal = {Nature},
  year    = {2020},
  volume  = {586},
  number  = {7828},
  pages   = {207--216},
  doi     = {10.1038/s41586-020-2764-0}
}

@article{Harris2018,
  author  = {Harris, Nicholas C. and Carolan, Jacques and Bunandar, Darius and Prabhu, Mihika and Hochberg, Michael and Baehr-Jones, Tom and Fanto, Michael L. and Smith, Alexander M. and Tison, Christopher C. and Alsing, Paul M. and Englund, Dirk},
  title   = {{Linear programmable nanophotonic processors}},
  journal = {Optica},
  year    = {2018},
  volume  = {5},
  number  = {12},
  pages   = {1623--1631},
  doi     = {10.1364/OPTICA.5.001623}
}

@article{Zhou2022,
  title = {{Photonic Matrix Multiplication Lights up Photonic Accelerator and Beyond}},
  author = {Zhou, Hailong and Dong, Jianji and Cheng, Junwei and Dong, Wenchan and Huang, Chaoran and Shen, Yichen and Zhang, Qiming and Gu, Min and Qian, Chao and Chen, Hongsheng and Ruan, Zhichao and Zhang, Xinliang},
  year = 2022,
  month = feb,
  journal = {Light: Science \& Applications},
  volume = {11},
  number = {1},
  pages = {30},
  issn = {2047-7538},
  doi = {10.1038/s41377-022-00717-8},
  url = {https://www.nature.com/articles/s41377-022-00717-8},
  urldate = {2026-05-24},
}

@article{Miller2013,
  author  = {Miller, David A. B.},
  title   = {{Self-configuring universal linear optical component}},
  journal = {Photonics Research},
  year    = {2013},
  volume  = {1},
  number  = {1},
  pages   = {1--15},
  doi     = {10.1364/PRJ.1.000001}
}

@article{FSLW_chip,
    author = {Ilya Kondratyev and Veronika Ivanova and Suren Fldzhyan and Artem Argenchiev and Nikita Kostyuchenko and Sergey Zhuravitskii and Nikolay Skryabin and Ivan Dyakonov and Mikhail Saygin and Stanislav Straupe and Alexander Korneev and Sergei Kulik},
    journal = {Photon. Res.},
    number = {3},
    pages = {A28--A40},
    publisher = {Optica Publishing Group},
    title = {Large-scale error-tolerant programmable interferometer fabricated by femtosecond laser writing},
    volume = {12},
    month = {Mar},
    year = {2024},
    url = {https://opg.optica.org/prj/abstract.cfm?URI=prj-12-3-A28},
    doi = {10.1364/PRJ.504588},
}

@misc{FSLW_Milan,
      title={A low-loss, 24-mode laser-written universal photonic processor in a glass-based platform}, 
      author={Andrea Barzaghi and Maëlle Bénéfice and Francesco Ceccarelli and Giacomo Corrielli and Valerio Galli and Marco Gardina and Vittorio Grimaldi and Jakub Kaczorowski and Francesco Malaspina and Roberto Osellame and Ciro Pentangelo and Andrea Rocchetto and Alessandro Rudi},
      year={2025},
      eprint={2505.01609},
      archivePrefix={arXiv},
      primaryClass={quant-ph},
}

@misc{fldzhyan2026nativeqrfactorizationprogrammable,
      title={{Native QR Factorization on Programmable Photonic Meshes}}, 
      author={S. A. Fldzhyan and S. S. Straupe and M. {\relax Yu}. Saygin},
      year={2026},
      eprint={2602.20701},
      archivePrefix={arXiv},
      primaryClass={physics.optics}
}

@article{tangWaveguidemultiplexedPhotonicMatrix2025,
  title = {Waveguide-Multiplexed Photonic Matrix--Vector Multiplication Processor Using Multiport Photodetectors},
  author = {Tang, Rui and Okano, Makoto and Zhang, Chao and Toprasertpong, Kasidit and Takagi, Shinichi and Takenaka, Mitsuru},
  year = 2025,
  month = jun,
  journal = {Optica},
  volume = {12},
  number = {6},
  pages = {812},
  issn = {2334-2536},
  doi = {10.1364/OPTICA.552023},
  url = {https://opg.optica.org/abstract.cfm?URI=optica-12-6-812},
  urldate = {2025-12-28}
}

@article{yuHeavyTailsPruning2023,
  title = {Heavy Tails and Pruning in Programmable Photonic Circuits for Universal Unitaries},
  author = {Yu, Sunkyu and Park, Namkyoo},
  year = 2023,
  month = apr,
  journal = {Nature Communications},
  volume = {14},
  number = {1},
  pages = {1853},
  issn = {2041-1723},
  doi = {10.1038/s41467-023-37611-9},
  url = {https://www.nature.com/articles/s41467-023-37611-9},
  urldate = {2026-06-03}
}

@article{chaiTimeWavelengthmultiplexedPhotonic2026,
  title = {Time- and Wavelength-Multiplexed Photonic Matrix-Matrix Multiplication Processor with on-Chip Wavelength (de)Multiplexers},
  author = {Chai, Chengli and Tang, Rui and Okano, Makoto and Toprasertpong, Kasidit and Takenaka, Mitsuru},
  year = 2026,
  month = may,
  journal = {Optics Express},
  volume = {34},
  number = {10},
  pages = {17539},
  issn = {1094-4087},
  doi = {10.1364/OE.593081},
  url = {https://opg.optica.org/abstract.cfm?URI=oe-34-10-17539},
  urldate = {2026-06-03}
}

@article{Taballione_2021,
    doi = {10.1088/2633-4356/ac168c},
    url = {https://doi.org/10.1088/2633-4356/ac168c},
    year = {2021},
    month = {aug},
    publisher = {IOP Publishing},
    volume = {1},
    number = {3},
    pages = {035002},
    author = {Taballione, Caterina and van der Meer, Reinier and Snijders, Henk J and Hooijschuur, Peter and Epping, Jörn P and de Goede, Michiel and Kassenberg, Ben and Venderbosch, Pim and Toebes, Chris and van den Vlekkert, Hans and Pinkse, Pepijn W H and Renema, Jelmer J},
    title = {A universal fully reconfigurable 12-mode quantum photonic processor},
    journal = {Materials for Quantum Technology},
}

@article{PIC_NNs,
    author = {Wang, Xinyi and Liao, Kun and Hu, Xiaoyong},
    title = {Algorithms, architectures, and platform implementations of integrated photonic neural networks},
    journal = {Applied Physics Reviews},
    volume = {13},
    number = {1},
    pages = {011327},
    year = {2026},
    month = {03},
    issn = {1931-9401},
    doi = {10.1063/5.0304866},
    url = {https://doi.org/10.1063/5.0304866},
}

@article{Reck1994,
  author  = {Reck, Michael and Zeilinger, Anton and Bernstein, Herbert J. and Bertani, Philip},
  title   = {{Experimental realization of any discrete unitary operator}},
  journal = {Physical Review Letters},
  year    = {1994},
  volume  = {73},
  number  = {1},
  pages   = {58--61},
  doi     = {10.1103/PhysRevLett.73.58}
}

@article{Carolan2015,
  author  = {Carolan, Jacques and Harrold, Christopher and Sparrow, Chris and Mart{\'\i}n-L{\'o}pez, Enrique and Russell, Nicholas J. and Silverstone, Joshua W. and Shadbolt, Peter J. and Matsuda, Nobuyuki and Oguma, Manabu and Itoh, Mikitaka and Marshall, Graham D. and Thompson, Mark G. and Matthews, Jonathan C. F. and Hashimoto, Toshikazu and O'Brien, Jeremy L. and Laing, Anthony},
  title   = {{Universal linear optics}},
  journal = {Science},
  year    = {2015},
  volume  = {349},
  number  = {6249},
  pages   = {711--716},
  doi     = {10.1126/science.aab3642}
}

@article{Harris2014,
  author  = {Harris, Nicholas C. and Ma, Yangjin and Mower, Jacob and Baehr-Jones, Tom and Englund, Dirk and Hochberg, Michael and Galland, Christophe},
  title   = {{Efficient, compact and low loss thermo-optic phase shifter in silicon}},
  journal = {Optics Express},
  year    = {2014},
  volume  = {22},
  number  = {9},
  pages   = {10487--10493},
  doi     = {10.1364/OE.22.010487}
}

@article{Qiu2020,
  author  = {Qiu, Huaqing and Liu, Yong and Luan, Chao and Kong, Dawei and Guan, Xiaowei and Ding, Yunhong and Hu, Hao},
  title   = {{Energy-efficient thermo-optic silicon phase shifter with well-balanced overall performance}},
  journal = {Optics Letters},
  year    = {2020},
  volume  = {45},
  number  = {17},
  pages   = {4806--4809},
  doi     = {10.1364/OL.400230}
}

@article{Watts2013,
  author  = {Watts, Michael R. and Sun, Jie and DeRose, Christopher and Trotter, Douglas C. and Young, Ralph W. and Nielson, Gregory N.},
  title   = {{Adiabatic thermo-optic Mach--Zehnder switch}},
  journal = {Optics Letters},
  year    = {2013},
  volume  = {38},
  number  = {5},
  pages   = {733--735},
  doi     = {10.1364/OL.38.000733}
}

@article{Wang2018,
  author  = {Wang, Cheng and Zhang, Mian and Chen, Xi and Bertrand, Maxime and Shams-Ansari, Amirhassan and Chandrasekhar, Sethumadhavan and Winzer, Peter and Lon{\v{c}}ar, Marko},
  title   = {{Integrated lithium niobate electro-optic modulators operating at {CMOS}-compatible voltages}},
  journal = {Nature},
  year    = {2018},
  volume  = {562},
  number  = {7725},
  pages   = {101--104},
  doi     = {10.1038/s41586-018-0551-y}
}

@article{Qi2020,
  author  = {Qi, Yifan and Li, Yang},
  title   = {{Integrated lithium niobate photonics}},
  journal = {Nanophotonics},
  year    = {2020},
  volume  = {9},
  number  = {6},
  pages   = {1287--1320},
  doi     = {10.1515/nanoph-2020-0013}
}

@article{PZT_cavity,
    author = {Chenlei Li and Tao Shu and Yueyang Zhang and Cunyu Shi and Wei Chen and Jianghao He and Fei Huang and Lijia Song and Zejie Yu and Ming Zhang and Yaocheng Shi and Daoxin Dai},
    journal = {Optica},
    number = {1},
    pages = {83--92},
    publisher = {Optica Publishing Group},
    title = {{Versatile wavelength-selective PZT photonic chips}},
    volume = {13},
    month = {Jan},
    year = {2026},
    url = {https://opg.optica.org/optica/abstract.cfm?URI=optica-13-1-83},
    doi = {10.1364/OPTICA.581491},
}

@misc{CompactLNOImodulator,
      title={{Enabling High-Bandwidth Coherent Modulation Through Scalable Lithium Niobate Resonant Devices}}, 
      author={Sadra Rahimi Kari and Paolo Pintus and John E. Bowers and Matt Robbins and Nathan Youngblood},
      year={2025},
      eprint={2502.10846},
      archivePrefix={arXiv},
      primaryClass={physics.optics},
}

@inproceedings{AbsorbsionModulator,
    author = {S. A. Srinivasan and P. Verheyen and R. Loo and I. De Wolf and M. Pantouvaki and G. Lepage and S. Balakrishnan and W. Vanherle and P. Absil and J. Van Campenhout},
    booktitle = {Optical Fiber Communication Conference},
    journal = {Optical Fiber Communication Conference},
    pages = {Tu3D.7},
    publisher = {Optica Publishing Group},
    title = {{50Gb/s C-band GeSi Waveguide Electro-Absorption Modulator}},
    year = {2016},
    url = {https://opg.optica.org/abstract.cfm?URI=OFC-2016-Tu3D.7},
    doi = {10.1364/OFC.2016.Tu3D.7},
}

@article{SwitchesReview,
    author = {Chen, Xiaojun and Lin, Jiao and Wang, Ke},
    title = {{A Review of Silicon-Based Integrated Optical Switches}},
    journal = {Laser \& Photonics Reviews},
    volume = {17},
    number = {4},
    pages = {2200571},
    doi = {https://doi.org/10.1002/lpor.202200571},
    url = {https://onlinelibrary.wiley.com/doi/abs/10.1002/lpor.202200571},
    year = {2023}
}

@misc{LowDepthPhotonMNIST,
      title={{Scalable optical neural network with nonlocally coupled coherent photonic processor}}, 
      author={Chun Ren and Ryota Tanomura and Kazuki Ichinose and Keigo Mizukami and Yoshitaka Taguchi and Taichiro Fukui and Yoshiaki Nakano and Takuo Tanemura},
      year={2026},
      eprint={2603.07174},
      archivePrefix={arXiv},
      primaryClass={physics.optics},
}

\end{document}